\documentclass[fleqn,usenatbib]{mnras}
\usepackage{newtxtext,newtxmath}
\usepackage[T1]{fontenc}
\usepackage{ae,aecompl}
\usepackage{natbib}

\usepackage{graphicx}	
\usepackage{amsmath}	
\graphicspath{{./}{figures/}}
\usepackage{xcolor}
\usepackage{rotating}
\usepackage[normalem]{ulem}
\usepackage{lipsum}  
\usepackage{soul}
\usepackage{graphicx}
\usepackage{hyperref}
\hypersetup{
  colorlinks   = true, 
  urlcolor     = blue, 
  linkcolor    = blue, 
  citecolor   = blue 
}

\newcommand{\orcid}[1]{\textsuperscript{\href{http://orcid.org/#1}{
\hskip2pt\includegraphics[width=8pt]{Orcid-ID.png}}}}

\title[Mapping the spectral index of Cas~A]{
Mapping the spectral index of Cas A: \\
evidence for flattening from radio to infrared
}

\author[V.~Dom{\v c}ek et al.]{
V.~Dom{\v c}ek,$^{1,2}$\thanks{E-mail: v.domcek@uva.nl}
J.~Vink,$^{1,2,3}$
J.~V. Hern\'andez Santisteban,${^4}$
T.~DeLaney,${^5}$
and P.~Zhou${^1}$ 
\\
$^{1}$Anton Pannekoek Institute for Astronomy, University of Amsterdam, Science Park 904, 1098 XH Amsterdam, The Netherlands\\
$^{2}$GRAPPA, University of Amsterdam, Science Park 904, 1098 XH Amsterdam, The Netherlands\\
$^{3}$SRON, Netherlands Institute for Space Research, Utrecht, The Netherlands\\
$^{4}$SUPA, Physics and Astronomy, University of St Andrews, St. Andrews, KY16 9SS, Scotland, UK \\
$^{5}$Physics and Engineering Department, West Virginia Wesleyan College, Buckhannon, WV 26201, USA 
}

\date{Accepted XXX. Received YYY; in original form ZZZ}

\pubyear{2020}

\begin{document}
\label{firstpage}
\pagerange{\pageref{firstpage}--\pageref{lastpage}}
\maketitle

\begin{abstract}
Synchrotron radiation from supernova remnants are caused by electrons accelerated through diffusive shock acceleration (DSA). The standard DSA theory predicts an electron spectral index of $p=2$, corresponding to a radio spectral index of $\alpha=-0.5$. An extension of DSA predicts that the accelerated particles changes the shock structure, resulting in a spectrum that is steeper than $p>2$ ($\alpha<-0.5$) at low energies and flattens with energy. 
For Cas A, a synchrotron spectral flattening was previously reported for a small part of the remnant in the mid-infrared regime. Here, we present new measurements for spectral flattening using archival radio (4.72~GHz) and mid-infrared (3.6~$\mu$m) data, and produce a complete spectral index map to investigate the spatial variations within the remnant.
We compare this to measurements of the radio spectral index from L-(1.285~GHz) and C-(4.64~GHz) band maps.
Our result shows overall spectral flattening across the remnant ($\alpha _\mathrm{R-IR} \sim -0.5$ to $-0.7$), to be compared to the radio spectral index of $\alpha_{\rm R}=-0.77$.
The flattest values coincide with the locations of most recent particle acceleration. In addition to overall flattening, we detect a relatively steeper region in the southeast of the remnant ($\alpha _\mathrm{R-IR} \sim -0.67$). We explore whether these locally steeper spectra could be the result of synchrotron cooling,
which provides constraints on the local magnetic-field strengths/age of the plasma, suggesting $B\lesssim 2$~mG for an age of 100~yr, and even
$B\lesssim 1$~mG using the age of Cas A, in agreement with other estimates. 
\end{abstract}



\section{Introduction}

Cosmic rays (CRs) are highly energetic particles that have been discovered more than a century ago \citep{Hess1912} that span in the energy range from a few hundred MeV up to $10^{20}$~eV. Their spectral distribution up to $3\times 10^{15}$~eV is roughly a power
law with a negative index of $2.7$. The spectral break around $3\times 10^{15}$~eV is known as the ``knee".  Up to the knee the
CR spectrum is dominated by protons, and the CRs must be of Galactic origin \citep[e.g.][]{Hillas2005}. 
For a long time it has been suspected that the energy needed to accelerate these Galactic CRs are powered by supernovae \citep{Baade1934}. The later identification of radio-synchrotron emission from accelerated electrons in supernova remnants (SNRs) suggested that CR acceleration was taking place in SNRs rather than during the explosion itself \citep[e.g.][]{Ginzburg1964}.

The relativistic CR electrons produce radio synchrotron
emission \citep[see][ for a review]{Dubner2015} and some young SNRs even produce X-ray synchrotron radiation near their shock fronts \citep[][ for reviews]{Reynolds2008,Helder2012}.
The gamma-ray detection of many SNRs have confirmed the presence of CRs.
The gamma rays could either be caused by the same electron population that causes synchrotron emission, or arise from hadronic CRs, which produce gamma rays through interactions with the background gas. The latter has been positively identified
in at least a few SNRs \citep[e.g.][]{Ackermann2013}. There is not yet evidence that SNRs can accelerate up to the ``knee", but some
SNRs can produce CRs close to that energy. For example, the gamma-ray bright source RX J1713.7-3946 produces gamma rays up to 100~TeV \citep{Aharonian2007}.

The CR acceleration in SNRs is thought to be caused by diffusive shock acceleration (DSA), also known as  first-order Fermi acceleration  \citep{Krymskii1977,Axford1977,Bell1978,Blandford1978}. According to the DSA theory, 
charged particles can repeatedly scatter across the shock by the magnetic field irregularities present on both sides of the shock, thereby gaining energy. After each
cycle a small fraction of the particles will not return to the shock again.
The combination of these two effects will result in a power-law  
distribution with a (negative) spectral index of $p=2$ for the expected compression ratio $\chi = 4$ between shocked and unshocked medium. 
It was soon realised that if CR acceleration is efficient, the presence of CRs ahead of the shock will influence the unshocked plasma: the CRs will push against the unshocked plasma, compressing
it and setting it in motion \citep{Eichler1979,Ellison1984}. Moreover, escape of the highest energy CRs far ahead of the shock, will drain energy from the system. The net result is that the combined compression of the CR precursor and the shock compression itself will be larger than $\chi=4$, whereas the shock compression itself may be reduced as the flow will arrive at the shock with a slower speed, and, therefore, lower Mach number. These processes are addressed in the theory of non-linear DSA  \citep[][ for a review]{Malkov2001}. 

According to the theory of non-linear DSA the CR spectrum will be steeper than $p=2$ for the low energy particles, as a result of the lower shock compression, and it will be flatter than $p=2$ for the highest energy particles, as these sample the total compression ratio. The total compression ratio can be in most extreme forms of the theory larger than 10 \citep[e.g.][]{Baring1999,Ellison2000},
largely driven by very high CR acceleration efficiency of $>50\%$.
Indeed there is some evidence that the radio emission from young SNRs has the predicted deviation from power-law \citep[e.g. concave spectrum, ][]{Reynolds1992}. The asymptotic value for the spectral index is expected to be $p=1.5$ \citep{Malkov1997}.

The CRs in the precursor may also give rise to another non-linear effect: it may cause magnetic field amplification through the so-called Bell mechanism \citep{Bell2004}. 
X-ray synchrotron emission provides evidence for amplified magnetic field near SNR shocks  \citep{Vink2003,Helder2008}.

In recent years less attention has been paid to the non-linear DSA theory. First of all,
although there is some evidence for enhanced compression ratios \citep{Warren2005},
the enhancement seems to be modest \citep[see also][]{Ferrand2010}. Secondly, 
gamma-ray measurements indicate that the CR acceleration efficiency is not as high as 50\% or more, but of the order of 10\% \citep{Slane2014}, and for Cas A even as low as 3--6$\%$ of the kinetic energy \citep{Araya2010}.
Finally, the gamma-ray spectra of SNRs are more consistent with $p=2.2$ all the way up to the TeV regime \citep[e.g.][]{Ahnen2017}, instead of the predicted $p=1.5$.

Cas A is among the youngest known SNRs \cite[$\approx 340$ years,][]{Thorstensen2001} and the brightest SNR in the sky at radio wavelengths, and has for that reason played an important role
in the debate about the Galactic origin of CRs.
Its radio spectrum is rather steep compared to other young SNRs, $\alpha =-0.77$ {\citep[e.g.][]{Baars1977,Trotter2017}},
whereas other young remnants have $\alpha\approx -0.6$. The radio spectral index is related to the electron spectral index $p_e$ as $\alpha=(p_e-1)/2$. So for Cas A, $p_e=2.54$.
This suggests that the spectrum of Cas A has been potentially affected by non-linear DSA effects.
For comparison, the general SNR population has on average $\alpha\approx -0.5$ \citep{Green2019},
consistent with the $p=2$ predicted by DSA in the absence of non-linear effects \citep{Bell1978,Malkov2001}.

If the steep radio spectrum of Cas A is a signature of non-linear DSA, the
synchrotron spectrum is expected to flatten at very high frequencies. 
Indeed,  some  flattening is reported for the total source spectrum
at 247~GHz \citep{Mezger1986}, and  up to 550~Hz based on \textit{Planck} data \citep{Onic2015}. In addition, two studies of infrared data also reported that the synchrotron
spectrum could be flattening up to $10^{13}$~Hz \citep{Jones2003,Rho2003}.

Here we revisit the case for flattening of the Cas A spectrum, using archival \textit{Spitzer} data in the 3.6 $\mu$m, which is relatively free of warm dust emission. We combine this with a 4.72~GHz Very Large Array (VLA) radio map \citep{Thompson1980}, in order to obtain the average spectral index between the two wavelength bands.
Our aim was to specifically look for the spatial variation of the
spectral index at high frequencies. This is also of interest because
Cas A appears to accelerate electrons both at the forward and reverse shock, as indicated by the association of X-ray synchrotron filaments
with both shock regions \citep{Helder2008}. Moreover, the effects of non-linear DSA, i.e. the flattening spectrum, should at some point be reversed by the effect 
of the synchrotron energy losses of the highest energy electrons. In particular, a cooling (or age) break is expected where the cooling time of the electrons
equals the age of the source, resulting in an overall break of the spectrum of $\Delta \alpha=0.5$ \citep[e.g.][]{Longair2011}. The frequency of the cooling break
depends on the age and magnetic field of the plasma, which for Cas A is estimated to be around 0.5 mG, but in likely variations will be present
\citep{Rosenberg1970,Atoyan2000,Vink2003}. This puts the expected cooling break in the mid- to near-infrared regime, as we will discuss later in Sec.~\ref{sec:cooling_break}.

\section{Observations and data analysis}

\begin{figure*}
	\includegraphics[width=\linewidth]{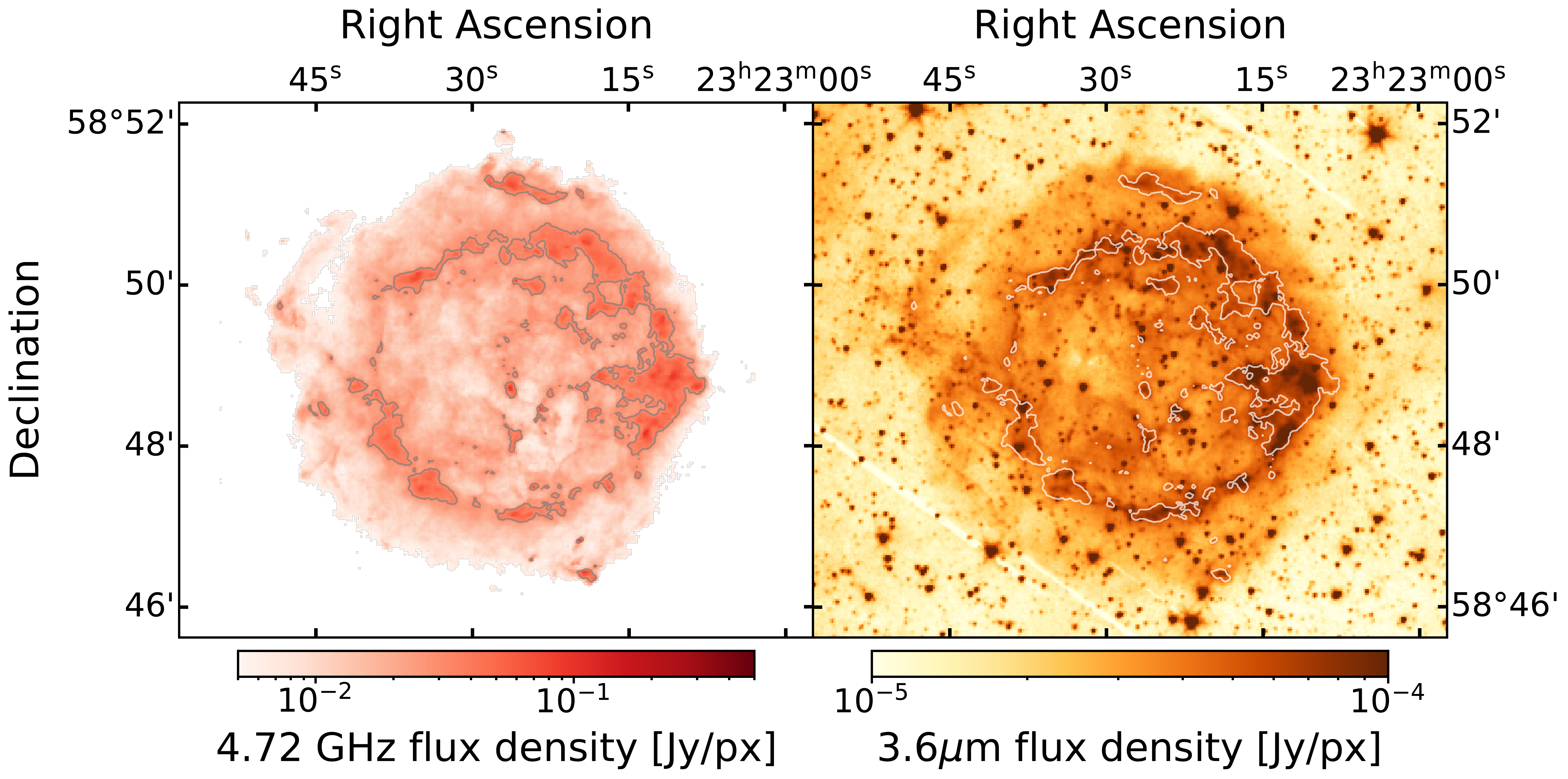}
    \caption{Flux density maps of Cas A. Left: 4.72 GHz image as obtained by VLA in 2001 \citep{Delaney2004} and flux calibrated to year 2009 (see section \ref{sec:radio_corrections} for more details). Right: \textit{Spitzer} 3.6 $\mu$m image as observed by PI:Arendt in 2009. Radio contours with flux density 0.03~Jy~px$^{-1}$ are displayed.}
    \label{fig:data}
\end{figure*}

\subsection{Infrared} 
Infrared data were observed in August 2009 (ObsID 34836224; PI: R.G.{Arendt}) using the \textit{Spitzer Infrared Array Camera} (IRAC) \citep{Fazio2004} and obtained from the \textit{Spitzer} Heritage Archive. We used the mosaiced products of the extended pipeline (level 2) that combines individual calibrated frames with the refined telescope pointing. {For reasons explained at the end of this section we restrict ourselves to the channel 1 (3.6~$\mu $m) data, which have a spatial resolution of 2-2.5$^{\prime\prime}$ \citep{Ennis2006}. These data contain the flux measurements (Fig.~\ref{fig:data}) and their uncertainties, both with pixel sizes of 0.6$^{\prime\prime}$}. 

\subsubsection{Infrared corrections and the Monte Carlo uncertainty propagation} 
\label{sec_corrections}
The infrared data require several corrections to be meaningfully used in the analysis. 
We correct for {\it a)} infrared environmental background, {\it b)} point-source contribution from the stars, {\it c)} extinction in the line of sight, and {\it d)} contributions of the other sources of emission.
Each correction of our analysis introduces additional uncertainties. We show the individual contributing steps in the Fig.~\ref{fig:error_diagram}.

To tackle all of these contributions, their uncertainties, and inter-dependencies we employ {Monte Carlo} simulations to
{propagate the uncertainties and assess their effects on our final spectral index map product.}
This method has a few advantages over using the analytical/semi-analytical method, as it accounts for a skewness in the distributions and allows for a better propagation of uncertainties. We use 10 000 simulation points. For clarity we show the applied procedure for a single pixel in the appendix of this paper.

{The process starts with a random generation of a normal distribution of the infrared flux for each pixel based on the flux and uncertainty provided by the data. }

The first correction we consider is the infrared background, visible in the uncorrected map of the remnant (right panel of Fig.~\ref{fig:data}). We use regions in southwest and northwest of Cas A to estimate the level of the background, assumed to be uniform across the remnant.
We find that the average flux density per pixel is $\approx 1.00\times10^{-5}$, with a 1-$\sigma$ uncertainty of $0.15\times10^{-5}$~Jy~px$^{-1}$. Based on this we simulate a normal distribution {using the Monte Carlo method, which} we subtract from the {infrared flux distributions}. However, as Cas A is located in a rather complex environment (e.g. structures east and west of the remnant), we also test how increasing the estimated uncertainty to $0.30\times10^{-5}$~Jy~px$^{-1}$ affects the significance of our results.

\begin{figure}
	\includegraphics[width=\columnwidth]{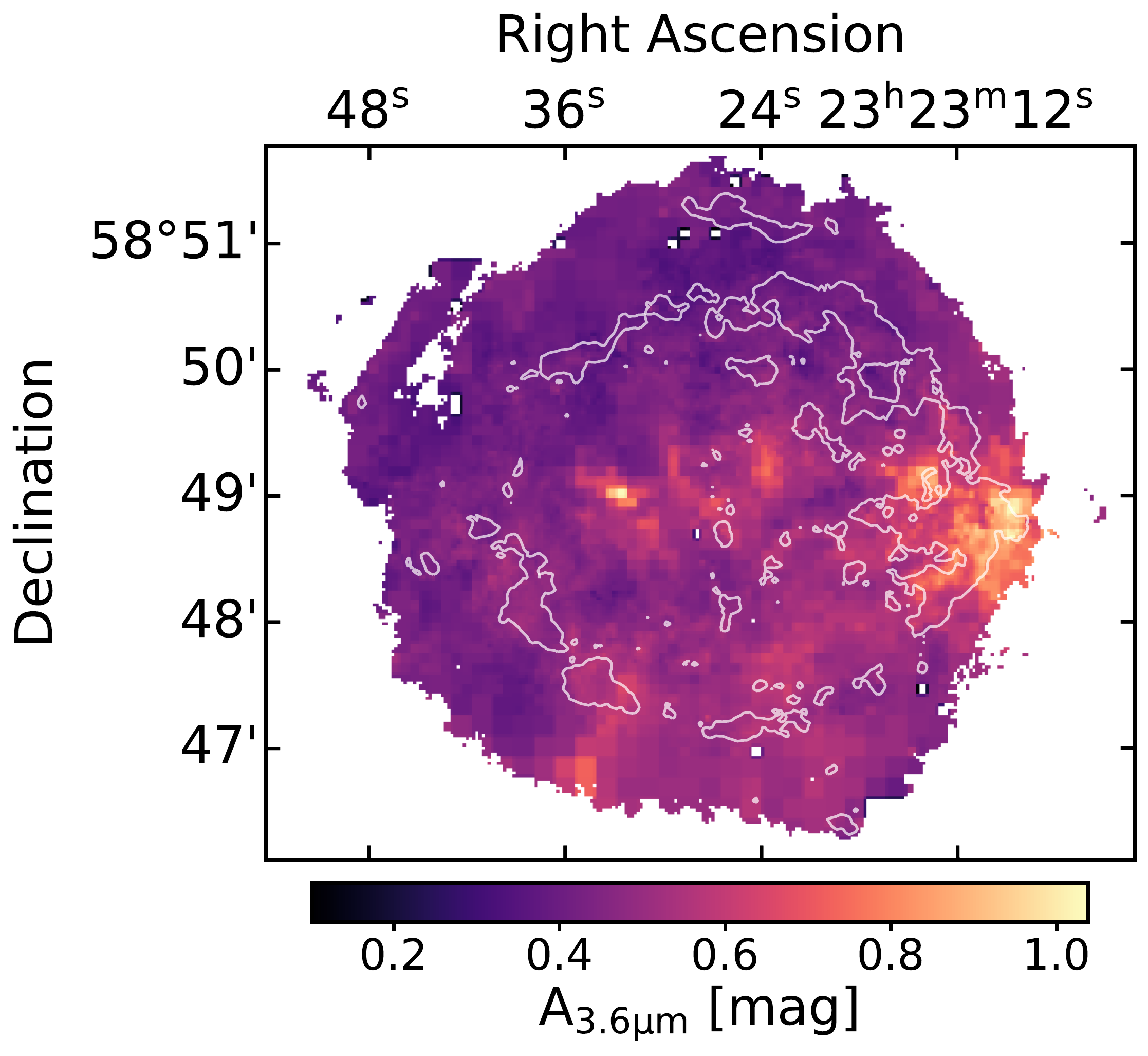}
    \caption{Extinction correction map at 3.6~$\mu$m based on X-ray N$_{\mathrm{H}}$ measurements \citep{Hwang2012}. }
    \label{fig:fig_NH_extintion_map}
\end{figure}

The second consideration in our method is taking into account the infrared extinction. \cite{Hurford1996} found extinction in Cas A to be varying between 4.6 $<$ $A_{\mathrm{v}}$ $<$ 6.2~mag increasing towards the west of the remnant. Recent X-ray study measuring the hydrogen column density \citep{Hwang2012} further confirm these results and using the \cite{Guver2009} $N_{\mathrm{H}}-A_{\mathrm{v}}$ relation places a lower limit of the extinction to $A_{\mathrm{v}} = 5$~mag. The western part of the remnant shows even stronger extinction with values going up to $A_{\mathrm{v}} = 15$~mag. We use $N_{\mathrm{H}}$ measurements of \citet[][]{Hwang2012} and assume a 10$\%$ fractional uncertainty in each pixel as this uncertainty was not provided. 
{Also in this case we generate normal distributions and subject the outcome to the $N_{\mathrm{H}} - A_{\mathrm{v}}$ relation of \citet[][ Eq.~1]{Guver2009} 
\begin{equation}
    N_\mathrm{H} (\mathrm{cm}^{-2}) = (2.21 \pm 0.09) \times 10^{21} A_\mathrm{v} (\mathrm{mag}).
\label{eq_guver}
\end{equation}
In order to further transform optical extinction ($A_\mathrm{v}$) into our required wavelength ($A_\mathrm{3.6\mu m}$), we relate the optical extinction to K-band extinction using $A_\mathrm{v}/A_\mathrm{K} = 8.8$ \citep[appropriate for the case of $R_\mathrm{v}$$ = 3.1$,][]{Cardelli1989}. 
For converting $A_\mathrm{K}$ to $A_\mathrm{3.6\mu\mathrm{m}}$ we use the extinction curve of \citet[][ Eq.~4]{Indebetouw2005} 
\begin{multline}
    \log(A_\lambda / A_K) = 0.61(\pm0.04) -2.22(\pm0.17) [\log(\lambda)] \\ +1.21(\pm0.23) [\log(\lambda)]^2
\label{eq_indebetouw}
\end{multline}
to obtain final extinction map (Fig.~\ref{fig:fig_NH_extintion_map}) and apply it to the \textit{Spitzer} dataset.}

Stars provide an additional infrared background source in the image. Their presence can increase the measured flux density, and locally change the measured spectral index. 
{Initially we tried to subtract the point sources using the {\sc daophot} extension of the {\sc iraf} data reduction software \citep{Tody:1986}. This approach was working reliably for the lower flux stars but turned out to be problematic in case of the bright and saturated ones, where the point source model could not reliably match the data.
Additionally, false detections in the filamentary shock structure caused alteration of non-affected regions (Fig.~\ref{figB:spitzer_star_sub}). For these reasons we decided for exclusion and masking of the brighter stars instead of their subtraction. We nevertheless still made use of the point source subtracted map for making of the lower resolution spectral index map in the section~\ref{sec_index_comparison}.

Masking was performed using \texttt{DAOStarFinder}\footnote{\url{https://photutils.readthedocs.io/en/stable/api/photutils.detection.DAOStarFinder.html}}. A few of the lower flux stars are still present in the final image but due to their size and low-flux nature they do not significantly affect our analysis.} 

Another source of emission that usually contributes to infrared wavelengths is
thermal emission from the dust particles.
Based on their size and chemical composition the radiation can be produced in different wavelengths covering different \textit{Spitzer} channels. Dust in Cas A consists mostly of silicate dust \citep{DeLooze2017} and, due to the silicon grain radiative properties, {its contribution below $\lambda \sim 7~\mu$m is suppressed considerably} (see Figure 10 of \cite{Draine2007} and Figure 3 of \cite{Rho2008}). 

{There are potential other emission features at wavelengths shorter than $7~\mu$m. Most noticeably, \cite{Ennis2006} suggested [Fe II] could be contributing to channel 3 of up to 50\% of its detected flux, and \cite{Rho2012} reported detection of molecular double-peaked CO line emission in the $\sim 4.2-5.0$~$\mu$m range, which is covered by \textit{Spitzer} channel 2. 
The energy range of channel 1 could also contain contribution of polycyclic aromatic hydrocarbons (PAHs), but as shown in Fig.~2 of \cite{Reach2006}, these are not relevant for the case of Cas A.
Channel 1 is therefore the only one that appears to be dominated by synchrotron radiation, and its synchrotron origin is supported by several previous studies \citep{Rho2003,Jones2003,Ennis2006}. We therefore choose only Channel 1 ($\lambda \simeq 3-4$~$\mu$m) of the \textit{Spitzer} telescope data for our analysis. 
For more in-depth discussion on the origin of the emission in each of the \textit{Spitzer} channels we refer the reader to \cite{Ennis2006}.}

\begin{figure}
	\includegraphics[width=\columnwidth]{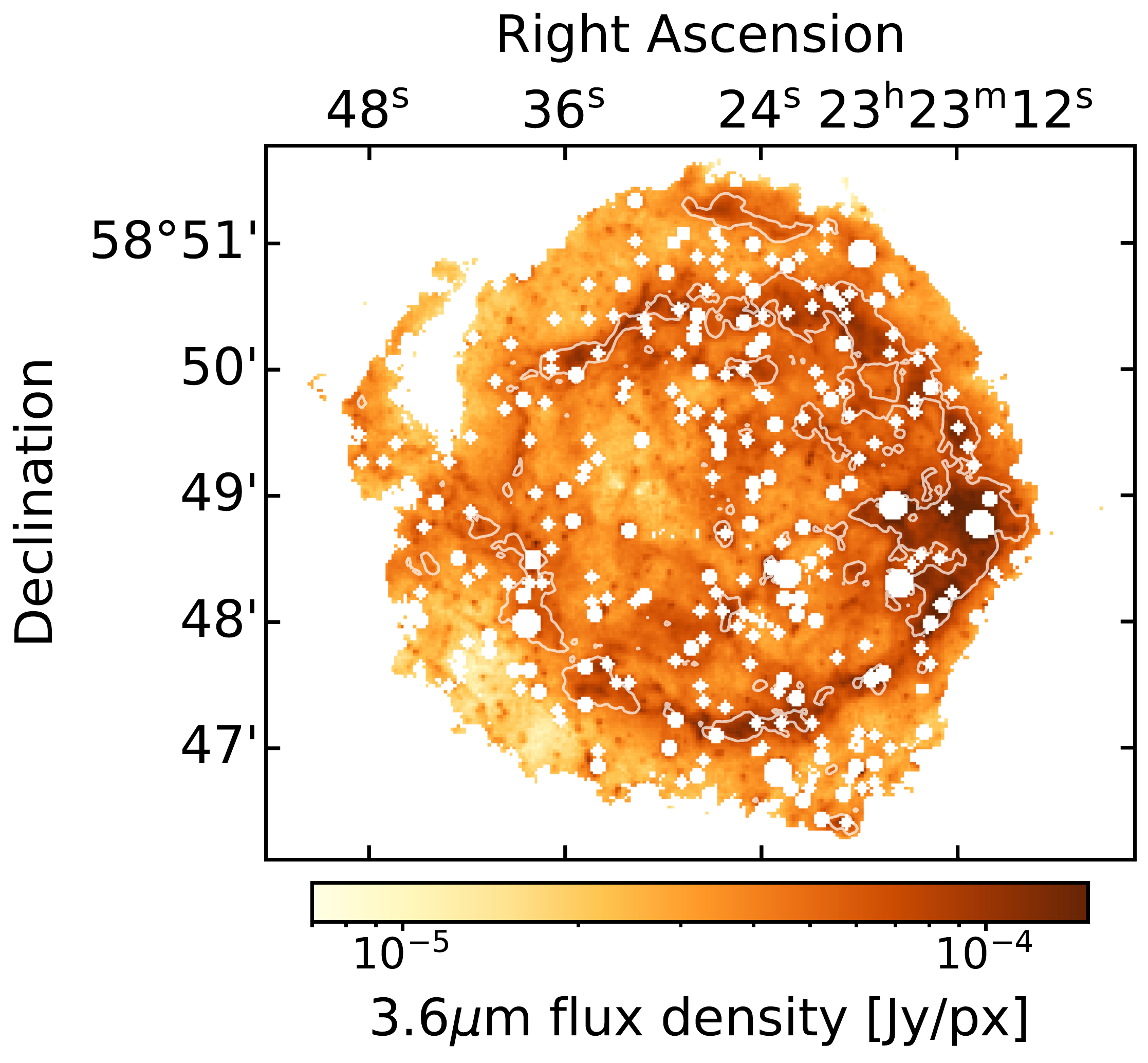}
    \caption{Spitzer 3.6~$\mu$m observation after background \& star subtraction and extinction correction. {Radio contour level $3 \times 10^{-2}$~Jy~px$^{-1}$ is displayed for easier orientation. } }
    \label{fig:fig_spitzer_corrected}
\end{figure}

\subsection{Radio}
\subsubsection{High resolution 4.72~GHz image}
\label{sec:radio_corrections}
For the construction of the high-resolution radio-to-infrared spectral index map we used a 4.72~GHz VLA (Very Large Array) radio image obtained in 2000-2001 \citep{Delaney2004}.
We performed the flux calibration by fitting a power-law of the flux density measurements from \cite{Perley2017} and obtaining the flux density at 4.72~GHz. For the total flux density to correspond with the year when \textit{Spitzer} data were obtained (2009), this value was further corrected for the secular fading using decline rate of 0.67\%~yr$^{-1}$, reported by \cite{Trotter2017}. The total flux density of Cas~A after these corrections is S$_{\mathrm{4.72 GHz}}$ = 697.6~Jy. Additionally we apply a low flux density mask on the data that masks out flux densities lower than {$5\times 10^{-3}$~Jy~px$^{-1}$} corresponding to the background level {and mask out regions that reach more than 10\% of the thermal noise contribution}. {The beam size of the observation is an ellipse with major axis 0.3792$^{\prime\prime}$, minor axis 0.3333$^{\prime\prime}$, position angle 90 degrees. The original map has an image pixel size of 0.1$^{\prime\prime}$}. We assume a 10$\%$ fractional uncertainty in the flux for each pixel as an upper limit of uncertainty of our radio data. 

\subsubsection{Additional radio data}
In order to have a reference radio spectral index map to investigate the flattening we also obtained the VLA L-band and C-band flux density images from \cite{DeLaney2014} where the data calibration and image preparation is described in full. The C-band image was normalised to 4.64~GHz and the L-band image was normalised to 1.285 GHz. The original images at 2.5$^{\prime\prime}$ resolution have a low dynamic range as reported in \cite{DeLaney2014}, which results in poor quality spectral index maps. Therefore, the flux density images were Gaussian smoothed to a resolution of 10$^{\prime\prime}$ to improve the signal-to-noise ratio for spectral index measurements. Since these flux density images are co-eval, there was no need to normalize them to the \cite{Perley2017} scale. The uncertainty in the integrated flux density at each band is about 2\% for the 1997-1998 observation epoch \citep{Perley1991}.

\begin{figure*}
	\includegraphics[trim=1cm 5cm 2cm 4cm,clip,width=0.85\linewidth]{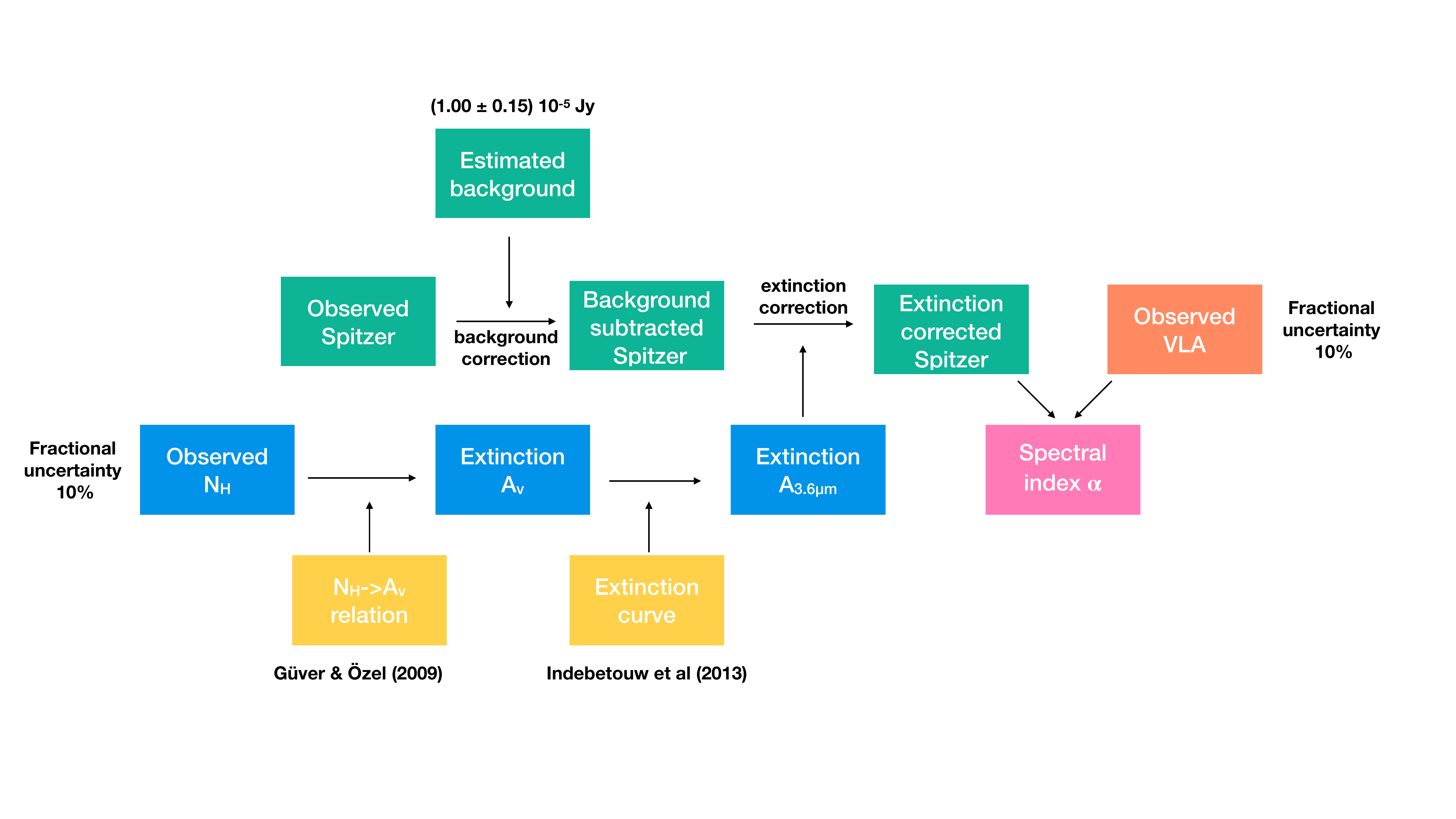}
    \caption{ Procedure towards obtaining the spatially-dependent spectral index $\alpha _\mathrm{R-IR}$. Each block represents a separate uncertainty that enters the analysis and is propagated using Monte Carlo simulations towards the final result shown in Fig.~\ref{fig:radio-to-ir_sp_index}. The colour-coding points to different data sets from which they arise and is defined as follows: \textit{Spitzer} data (green), Chandra data (blue), VLA data (orange), used equations in the process (yellow) and final spectral index (pink). This colour-coding is further used in the Appendix~\ref{sec:appendix} of the paper.} 
    \label{fig:error_diagram}
\end{figure*}

\subsection{Spectral index maps}
Creating spectral index maps is one of the methods to investigate the spectral evolution of extended objects like SNRs and the main method of this paper. The core of this method is in comparing flux measurements at two distinct frequencies. The further these frequencies are apart, the less dependent the measure spectral index is on systematic and statistical uncertainties of the flux measurements. In order to make the spectral index maps we first co-align radio and infrared images to the same coordinate grid with pixel size of $\approx$ 1.4$^{\prime\prime}$px$^{-1}$. We use the \texttt{reproject\_exact} tool in {\sc python}\footnote{\url{https://reproject.readthedocs.io/en/stable/api/reproject.reproject_exact.html}} to achieve this. 

\begin{figure*}
	\centering
	\includegraphics[width=\linewidth, angle=0]{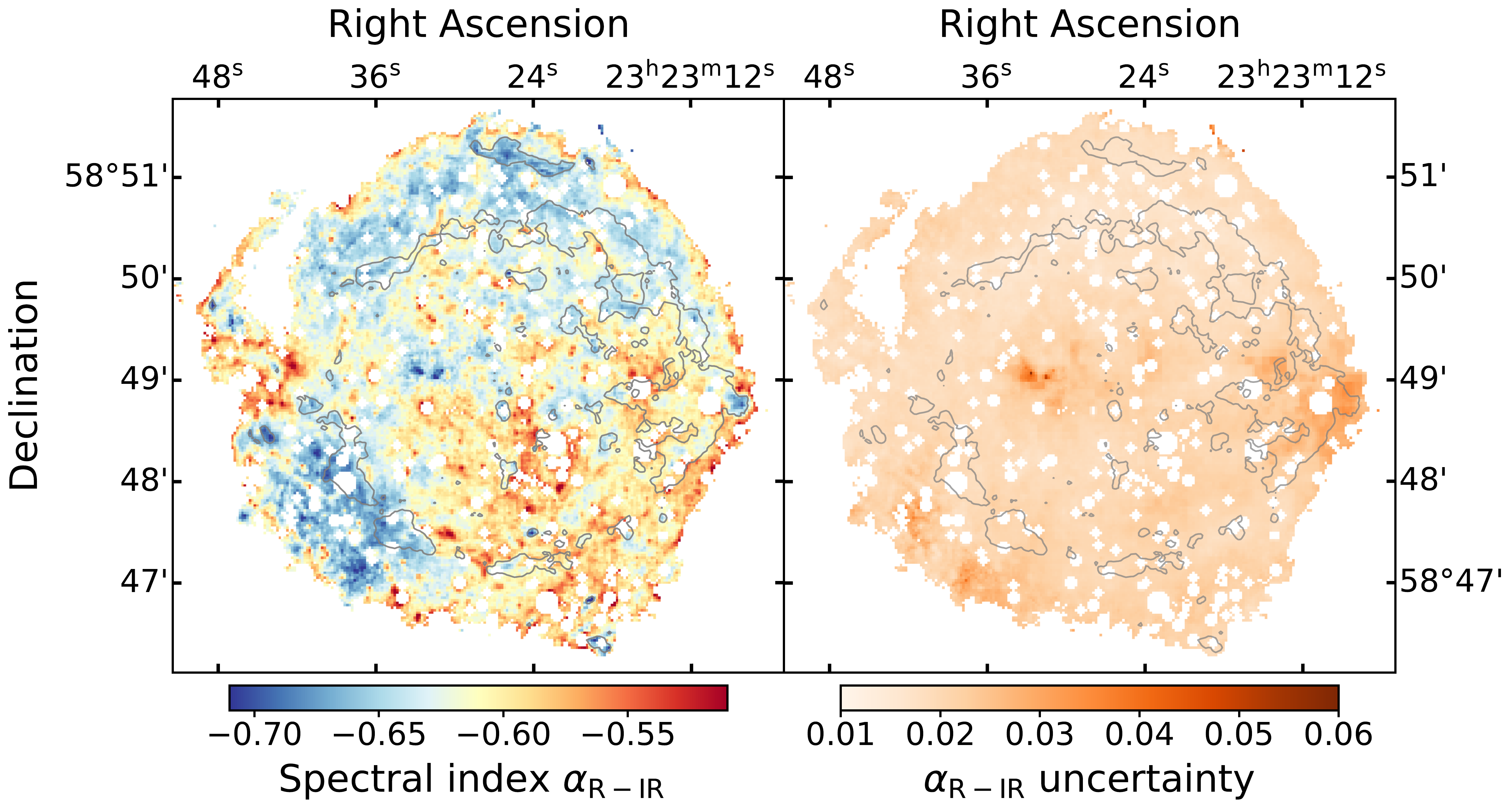}
	\caption{Radio to mid-infrared spectral index map ({left}) with its uncertainties ({right}). Radio contour level of $3 \times 10^{-2}$~Jy~px$^{-1}$ is displayed for easier orientation.}
    \label{fig:radio-to-ir_sp_index}
\end{figure*}

{We produce two spectral index maps using Eq.~\ref{eq:spectral_index}}
\begin{equation}
\alpha = \frac{\log({\mathrm{S}_{\nu _2}}) - \log{(\mathrm{S}_{\nu _1}})}{\log(\nu _2) - \log (\nu _1)}.
\label{eq:spectral_index}
\end{equation}
The radio-to-infrared spectral index map is constructed between the VLA (4.72~GHz) and the \textit{Spitzer} (3.6~$\mu$m) bands. $\mathrm{S}_{\nu _1}$ and $\mathrm{S}_{\nu _2}$ correspond to calibrated flux per pixel in the radio and the infrared, while $\nu _1$ and $\nu _2$ are the corresponding frequencies of the input maps.  
Note that $S_{\nu _2}$ was subjected to our Monte Carlo method for obtaining uncertainty distributions.
We collapse these results into two maps that show the median spectral index value and its 1-$\sigma$ uncertainty in Fig.~\ref{fig:radio-to-ir_sp_index}.

{The reference radio-radio spectral index map between the \textit{u-v} matched lower resolution 1.285~GHz and the 4.64~GHz flux density images was computed using the {\sc aips} task \texttt{COMB}. In order to ensure that the spectral index between the two radio images is robust, a number of techniques were employed as described in \cite{DeLaney2014}. The most important of which is using the same 4.64 GHz image as a default when using the {\sc aips} maximum entropy deconvolution routine \texttt{VTESS} to make the final 1.285~GHz and 4.64~GHz flux density images \citep{Cornwell1985}. For weak radio sources, calibration errors caused by, for example, slightly different \textit{u-v} coverage of the radio observations and slightly different primary beam attenuation corrections can lead to spurious spectral index measurements. However, Cas~A is exceptionally bright and therefore the minor differences in the primary beam have very little effect and the strict matching of \textit{u-v} coverage is not required to get reliable spectral index measurements. Although the nominal spatial resolution of the 1.285~GHz and the 4.64~GHz flux density images is 2.5$^{\prime\prime}$ based on \textit{u-v} coverage, The two flux density images were also Gaussian smoothed to 10$^{\prime\prime}$ resolution to improve the signal-to-noise ratio and allow reliable spectral index measurements even in fainter regions.}

\begin{figure*}
    \centering
    \includegraphics[width=\linewidth]{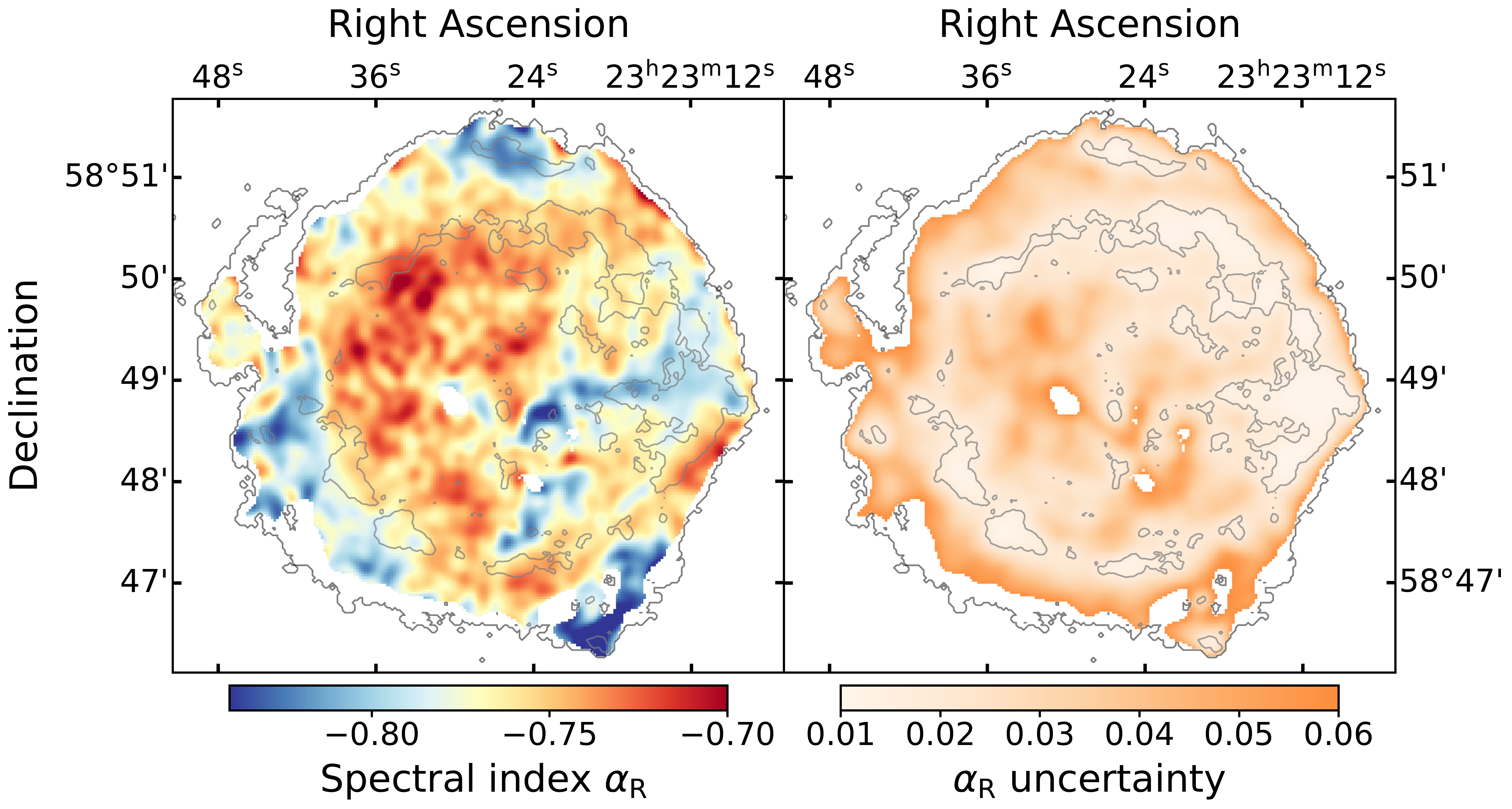}
    \caption{Radio spectral index between L-band (1.285~GHz) and C-band (4.64~GHz) based on 1997/1998 radio data of \citet{DeLaney2014} (left).
    Uncertainty of the radio spectral index (right). In addition to radio contours of flux density 0.03~Jy~px$^{-1}$ we also display the outer boundary of radio-to-infrared spectral index map from Fig.~\ref{fig:radio-to-ir_sp_index}.}
    \label{fig:radio_sp_index}
\end{figure*}

\begin{figure*}
    \centering
    \includegraphics[width=\linewidth]{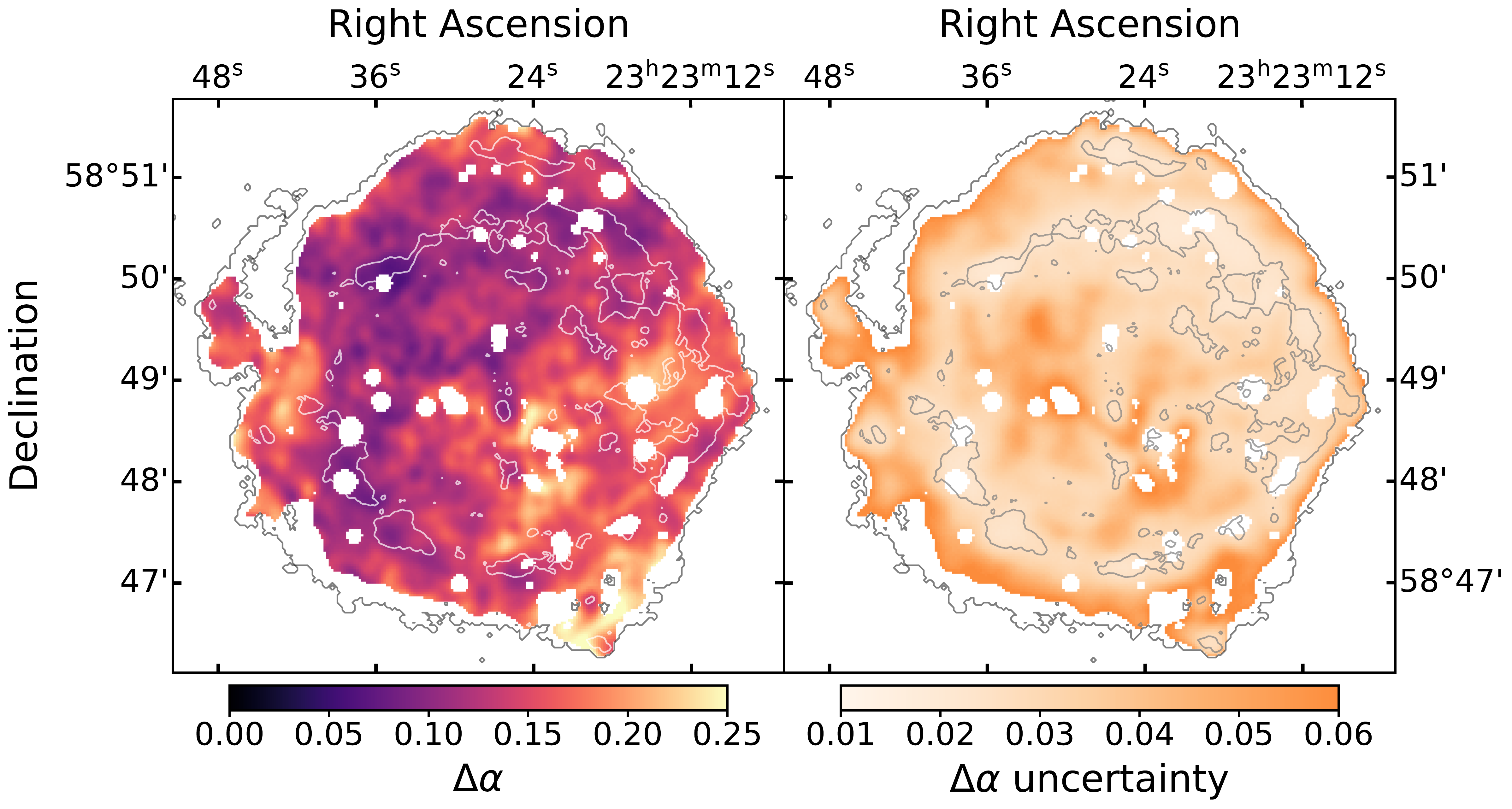}
    \caption{Deviation between radio spectral index ($\alpha _\mathrm{R}$) and radio-to-infrared spectral index ($\alpha _\mathrm{R-IR}$). In addition to radio contours of flux density 0.03~Jy~px$^{-1}$ we also display the outer boundary of radio-to-infrared spectral index map from Fig.~\ref{fig:radio-to-ir_sp_index}. 
    }
    \label{fig:diff_sp_index}
\end{figure*}

\section{Results}
\label{sec_Results}
We produced spectral index maps between the radio (4.72~GHz) and infrared (3.6~$\mu$m) (see Fig.~\ref{fig:radio-to-ir_sp_index}), {and between two radio frequencies (1.285 - 4.64 GHz)}. The results show an overall significant flattening of the average radio spectral index $\alpha _\mathrm{R} = -0.7653 \pm 0.0002$ to the radio-infrared index $\alpha _\mathrm{R-IR} = -0.6143 \pm 0.0002$. 

\subsection{Radio-to-infrared spectral index}
Fig.~\ref{fig:radio-to-ir_sp_index} shows that radio-to-infrared spectral index map has some structure to it, although in all cases the
spectral index is flatter than the average radio spectral index of $\alpha _\mathrm{R} \approx -0.77$.

The flattest index (with values of $\alpha _\mathrm{R-IR} \approx -0.60$) is found in the {outer rim} regions close to the forward {shock and over majority of the southwestern quadrant of the remnant}. The index steepens slightly to $\alpha _\mathrm{R-IR} \approx -0.62$ in-between the forward and reverse shocks, most visibly in the northern half of the remnant. The region that stands out most in the {radio-to-infrared} spectral index map is located in the southeastern part of the remnant and has a relatively steep spectral index of $\alpha _\mathrm{R-IR} \approx -0.66$. {This relatively steep spectral index is at least partially caused by the lack of emission in the mid-infrared band, that is also noticeable in the other \textit{Spitzer} channels (4.5, 5.8, 8.0 $\mu$m) and in the X-ray non-thermal emission \citep{Helder2008}. The available portion of this region in the radio spectral index map in Fig.~\ref{fig:radio_sp_index} also appears to be steeper, although the difference from the mean radio spectral index is smaller than in the mid-infrared case.}

In order to compare these interesting regions with the results for the whole remnant, we produce separate histograms for some regions and calculate their mean spectral index {in flux density space}. For a clear display of the results, we specify the regions as:
i) outer rim (pink colour), ii) northern arc (blue colour), iii) southeast (purple colour) and southwest (green colour). We show their locations in Fig.~\ref{fig:regions_mask} and list the mean values in Table~\ref{tab:sp_region_means}, and the respective histograms in Fig.~\ref{fig:hist_all} (left). 

{We tested whether we are not over-subtracting the background in the remnant by performing our analysis on one of the low-flux pixels in the southeast region. We find that even in this low-flux region, flux is still sufficient for the background correction to proceed and produce a Gaussian shaped distribution (see Fig.~\ref{figA:observed_data}}).

We further tested the effect of increasing the uncertainty of the background to $3\times 10^{-5}$~Jy~px$^{-1}$. Although it does not change the spectral index values by itself, it can in the most affected low mid-infrared flux regions increase the uncertainty up to $0.06$. This however still points towards significant enough spectral flattening.

\subsection{Radio spectral index}
The lower resolution radio spectral index map (Fig.~\ref{fig:radio_sp_index}) portrays a slightly different picture compared to radio-to-infrared spectral index. The mean radio spectral index for the whole remnant is {$0.7653 \pm 0.0002$} which is in agreement with the previously measured values of \cite{Baars1977} and \cite{Trotter2017}. However, in terms of the spatial distribution flatter indices appear mainly in the center east, turning into steeper values in the west. In-between the forward and reverse shock the situations appears to be similar to $\alpha _\mathrm{R-IR}$ where we observe close to average or slightly steeper values in locations we dubbed northern arc and southeast. Unfortunately due to the limits of the radio images we cannot reliably construct pixel-to-pixel spectral index in the outer rim region. We can nevertheless compute the mean spectral index from the radio flux density maps, which points towards similar steeper values as in-between the shocks. 

The tip of the southwest region shows steep radio spectral index values. This region corresponds to one of the locations where the supernova shock wave has overrun a slow moving optical knot \cite{Braun1987} and \citet{Anderson1991}, resulting in a bow shock. The interaction has caused rapid brightening in 1970s and consequent fading since 1980s. However, the radio-to-infrared spectral index does not show any noticeable steepening within the same region. It is possible that the steep index in radio comes from a different electron population that is superimposed on top of the underlying typical average region. 

\begin{figure}
    \centering
    \includegraphics[width=\linewidth]{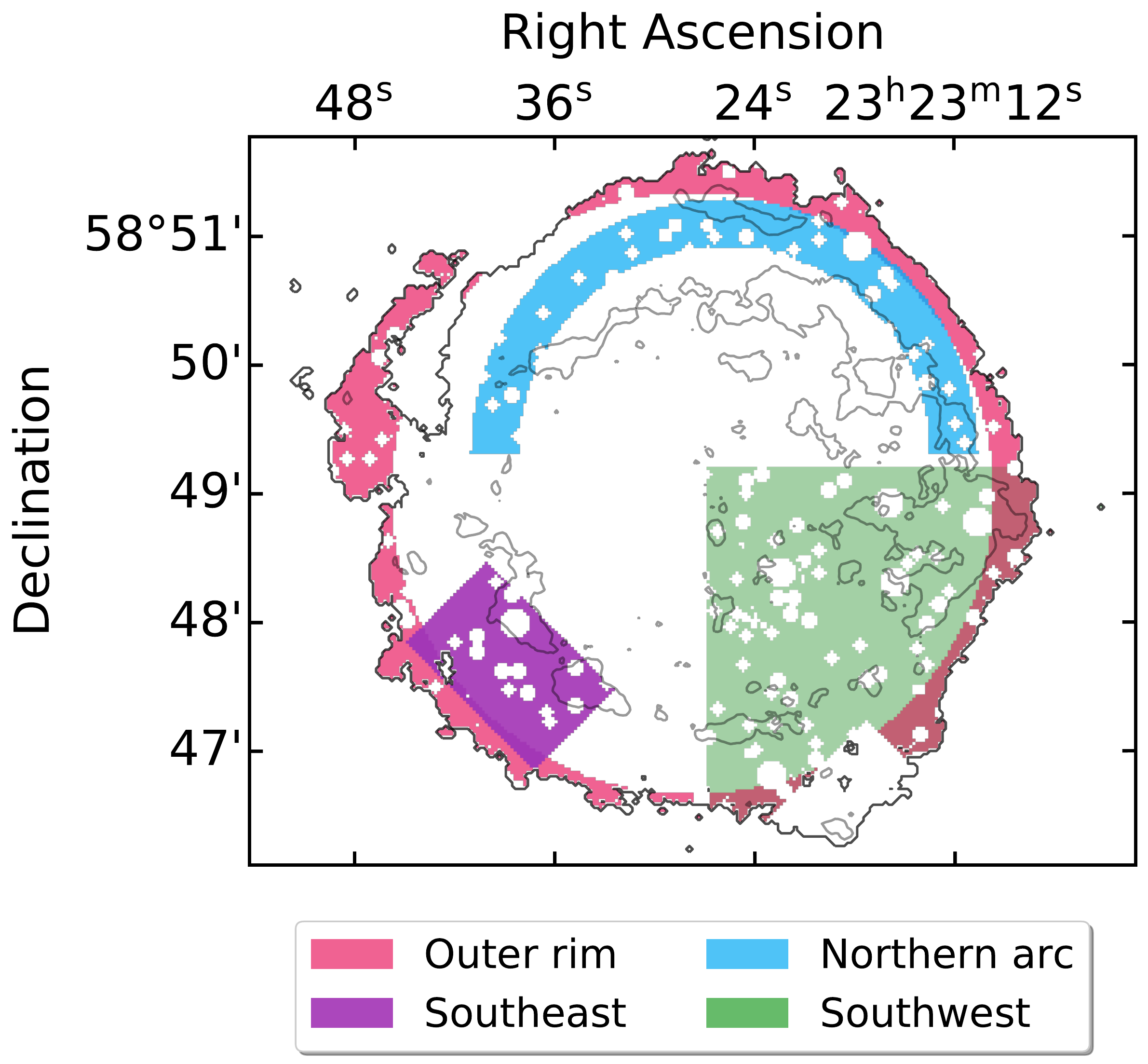}
    \caption{Masking regions used for regions of interest in Table~\ref{tab:sp_region_means}.}
    \label{fig:regions_mask}
\end{figure}

Table~\ref{tab:sp_region_means} summarises the mean radio spectral indices for all selected regions from Fig.~\ref{fig:regions_mask}.

\subsection{Spectral index comparison}
\label{sec_index_comparison}
{A comparison of the radio-to-infrared and inter-radio spectral index map can give a more complete picture of the spectral index evolution in the remnant and better point to regions of greater or lesser flattening. For that purpose we constructed a spectral index deviation map, $\Delta \alpha$ (Fig.~\ref{fig:diff_sp_index}). 

In order to produce the $\Delta \alpha$ map we required a radio-to-infrared spectral index map with spatial resolution of 10$^{\prime\prime}$. However, it was not possible to use Gaussian smoothing on the data from Fig.~\ref{fig:radio-to-ir_sp_index} due to the high number of masked out point sources in the remnant. 

We therefore took an alternative approach of producing radio-to-infrared spectral index map in the lower resolution. We circumvented the issue of the star masking by using a smoothed out point source subtracted map (Fig.~\ref{figB:spitzer_star_sub} and Fig.~\ref{figB:data_smoothed}). Although this approach may have resulted in a partial loss of flux in the filamentary structure due to false detection of point sources (the reason we don't use it in the higher resolution analysis), the consequent smoothing has largely filled out these over-subtracted locations and spread the possible deficiency of flux over larger region. 
A comparison of the two methods show that the impact on the measured $\alpha _\mathrm{R-IR}$ in table~\ref{tab:sp_region_means} does not deviate more than $\Delta \alpha \approx 0.005$ for any given averaged region.

Additionally as few saturated stars were not subtracted correctly, we masked out these locations in our lower-resolution spectral index map. We display the point source subtracted map, smoothed flux images and lower resolution spectral index map in the appendix of the paper (Fig.~\ref{figB:spitzer_star_sub}, \ref{figB:data_smoothed}, \ref{figB:radio-to-ir_sp_index_smoothed}).

The final 10$^{\prime\prime}$ spectral index deviation map (see Fig.~\ref{fig:diff_sp_index}) shows flattening everywhere by at least $\Delta \alpha \sim 0.05$. Three regions appear to be flattening significantly more ($\Delta \alpha > 0.15$) than in the rest of the remnant: {\it i)} the southwest, {\it ii)} the east --- a bit southwest to the base of the jet --- and {\it iii)} the north --- around the filamentary structure --- (see Fig.~\ref{fig:diff_sp_index}). 

The southwestern quadrant ({\it i}) is flattening more as a whole. This is a combination of for the most part relatively average index in radio and much flatter radio-to-infrared index. This quadrant has also stronger X-ray synchrotron continuum emission \citep[see Fig.~6 of][]{Helder2008}. 
The north location ({\it iii}) is centered around a bright filamentary structure that is steeper than average in both $\alpha _\mathrm{R}$ and $\alpha _\mathrm{R-IR}$, but still shows an above average flattening.
The east part ({\it ii}) is quite steep in the radio, but
flattens considerably towards infrared (Fig.~\ref{fig:diff_sp_index}).
It does not appear to have either the radio filamentary structure nor any strong X-ray synchrotron emission as in the previously pointed out location that could explain its flattening. This region 
contains some of the Fe-rich ejecta that has overtaken the silicon-rich
shell, and it does have a relatively high ionisation age
compared to other parts of the remnant \citep{Hwang2012}.
But it is not immediately clear what the connection is between these X-ray determined plasma properties and
the radio and radio-to-infrared spectral indices. 

In contrast, the southeastern region in the radio-to-infrared spectral index map does not seem to be standing out as much in Fig.~\ref{fig:diff_sp_index}. It experiences only mild flattening that is comparable to other regions. The region is nevertheless still interesting as together with the northern filament both of this locations have steeper than average spectra.}

\begin{figure*}
    \centering
    \includegraphics[width=\linewidth]{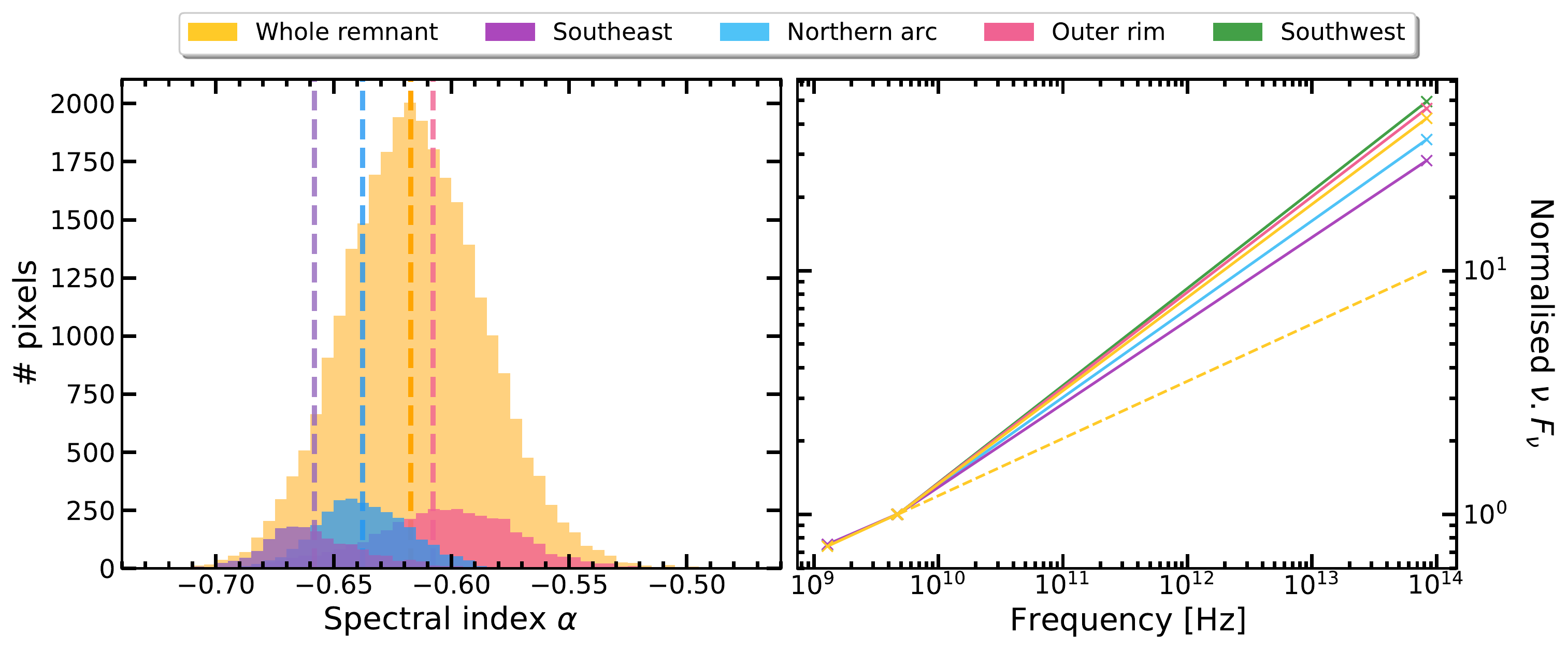}

    \caption{{Left:} Histogram of median spectral index values $\alpha _\mathrm{R-IR}$ for regions of interest. Outer rim region appears to be significantly flatter than the northern arc and southeast. 
    Dashed lines represent mean values of individual distributions.
    {{Right:} Spectral energy distribution for the same locations, including southwest region that has histogram spread similar to the outer rim. Yellow dashed line shows extrapolation of the mean radio spectral index for the whole remnant at higher frequencies. All regions have been normalised at 4.72~GHz.}
    }
    \label{fig:hist_all}
\end{figure*}

\begin{table*}
    \centering
    \begin{tabular}{l|c|c|c}
        \hline
         & Mean spectral index & Mean spectral index &  Deviation\\
        Region & $\alpha _\mathrm{R}$ & $\alpha _\mathrm{R-IR}$ &  $\Delta \alpha$\\
        \hline
        \hline
        Whole remnant & $-0.7652 \pm 0.0002$  & $-0.6173 \pm 0.0002$ & $0.1479 \pm 0.0003$  \\
        \hline 
        Outer rim & $-0.7777 \pm 0.0009$ & $-0.6078 \pm 0.0005$  & $0.170 \pm 0.001 $ \\
        Northern arc & $ -0.7742 \pm 0.0006$ & $ -0.6378 \pm 0.0004$  & $0.1364 \pm 0.0007$\\
        Southeast & $ -0.7795 \pm 0.0009$ & $ -0.6581 \pm 0.0006$ & $ 0.121 \pm 0.001$ \\
        Southwest & $ -0.7726 \pm 0.0004$ & $ -0.6011 \pm 0.0004$ & $ 0.1715 \pm 0.0005$ \\
        \hline
    \end{tabular}
    \caption{Mean spectral indices for regions of interest shown in Fig.~\ref{fig:hist_all}. $\alpha _\mathrm{R}$ represents radio spectral index while $\alpha _\mathrm{R-IR}$ is radio-to-infrared spectral index. $\Delta \alpha$ represents the difference between the two indices.}
    \label{tab:sp_region_means}
\end{table*}

\section{Discussion}

Our study shows that there is an overall flattening of the synchrotron spectrum of Cas A between the radio and infrared. Here we
compare our results to previous studies on the broad band spectral properties of the synchrotron emission, and discuss the results in the context
of non-linear shock acceleration models and a possible cooling break.

\subsection{Comparison with other studies}
\label{sec:disc_comparison_other_studie}

Evidence for spectral flattening of the synchrotron spectrum {with respect to the radio spectral index measured below 10~GHz} has been found before. High-frequency radio observations above 30 GHz already showed that the overall
spectrum of Cas A flattens at high frequencies \citep{Mezger1986,Onic2015}. This is also supported by infrared studies in the 2~$\mu$m range \citep{Jones2003,Rho2003}.
The most important difference between these studies and what is presented here is that the {high frequency} radio studies {explored} the spectral flattening of the whole
remnant, without the ability to detect spatial variation {across the whole remnant}. The infrared study by \cite{Jones2003} was restricted to small regions in the northwest part of the remnant,
and the {spectral tomography} study by \cite{Rho2003} was limited mainly to the bright shell, due to the low flux of diffuse emission in the K-band, and the relatively high background.

Moreover, we find on average, flatter spectral indices ($\alpha _\mathrm{R-IR} \sim -0.6$) compared to those reported at 2~$\mu$m ($\alpha _\mathrm{2\mu m} \sim -0.7$), even for comparable
locations. The discrepancy could be caused by either a general steepening at shorter wavelengths, or by differences in the employed methodology, for example by our more detailed treatment of extinction and uncertainties (see Fig.~\ref{fig:error_diagram}).

As there are some systematic uncertainties regarding the measurements of $N_{\rm H}$ of the order of 25\% \cite[see for example ][]{Zhou2018}, we tested the effect of a 25\% overestimation of $N_{\rm H}$ in our final spectral index. We found it to be $\Delta \alpha\approx 0.02$ at most, thus $N_{\rm H}$ overestimation cannot account for the discrepancy between our measurements and those using the emission at 2~${\rm\mu m}$.

A general steepening at shorter wavelength would be an interesting possibility, as it may hint at a synchrotron cooling break, similar to one discussed in Section~\ref{sec:cooling_break}.

The relativistic electron population responsible for the synchrotron emission will also produce gamma-ray emission through inverse Compton scattering, with a power law
slope that should be $\Gamma=\alpha+1$. However, for Cas A the gamma-ray emission is suspected to be dominated by hadronic gamma-ray radiation:
protons and other atomic nuclei colliding with background plasma, thereby producing neutral pions, which immediately decay into photons.
If we denote the particle spectral index with $p$, the gamma-ray spectral index for pion decay should be $\Gamma \approx p$, whereas for
inverse Compton scattering it should be $\Gamma= (p+1)/2$ (see \citet{Hinton2009} for a review on the gamma-ray emission from SNRs).

Several studies of the 
the gamma-ray spectrum in the GeV to TeV range 
covered by Fermi-LAT and the VERITAS and MAGIC
observatories all find a spectral index
for the protons of $p\approx 2.2$ \citep[e.g.][]{Yuan2013,Ahnen2017,Abeysekara2020}.
According to the theory of DSA, both electrons and hadrons are accelerated with the same spectral index in momentum. 
This also means that they should have the same spectral index in energy, provided the hadrons are relativistic, and the electron spectrum is not
affected by synchrotron cooling. So if the gamma-ray emission from Cas A is hadronic,
the accelerated electrons have the same slope, and we expect a radio spectral index of $\alpha= - (p-1)/2= -0.6$. 
This is indeed consistent with our results, and inconsistent with the spectral index below $\sim$10~GHz. 

\subsection{Consequences for the shock acceleration models}
\label{sec:disc_consequences_for_models}

The observed flattening of the spectrum from radio to infrared synchrotron radiation is in line with the predictions of non-linear DSA.
This theory predicts that the flattening is expected to occur in those electron populations accelerated by the most efficiently accelerating shocks, i.e those shock capable of
converting a large fraction of the shock energy into accelerated particles. Our results indicate that the flattest spectra are located in the southwestern quadrant. 
This is also the location of stronger non-thermal X-ray emission (see Fig.~6 of \cite{Helder2008}), where the forward and reverse shock are closest to each other. It is therefore probably a place of more efficient particle acceleration, which could exhibit stronger flattening effects as well. As discussed by \citet{Helder2008}, the western part is also the region where the reverse shock has the highest shock velocity in the frame of the ejecta.

Going from the forward shock inwards into the outer radio shell (northern arc region in Fig.~\ref{fig:regions_mask}) a gradual steepening of indices is observed in the $\alpha _\mathrm{R-IR}$ map. This is not as obvious in the $\Delta \alpha$ map, but this map has a lower spatial
resolution.
The steepening could indicate that in the past, when this electron population was accelerated,  the radio-to-infrared spectrum was steeper, and less affected by non-linear effects.
However, since the shock velocity was higher in the past, a more likely explanation is that the steeper spectum has been affected by synchrotron cooling. Moreover, the {\em radio} spectral index is typical for Cas A --- $\alpha_\mathrm{R}\approx -0.77$ --- which is the reason
why non-linear models need to be considered in the first place.

The flattest spectral indices that we measure near the shock regions 
are, however, not supporting the case for the extreme non-linear acceleration, which predicts particle spectral indices $p<2$ \citep{Malkov1997},
and hence synchrotron spectral indices  $\alpha <0.5$. 
In more recent updates of the non-linear DSA models \citep[e.g.][]{Vladimirov2008,Caprioli2009}, more emphasise is placed on the effects of magnetic-field amplification 
by the Bell mechanism \citep{Bell2004} on the energy distribution of the accelerated particles. In a recent paper, \citet{Bell2019} shows that the energy transferred from the accelerating particles to the magnetic field will also lead to steeper particle spectra.
Another way in which magnetic fields may affect the particle spectra is if the particles scatter off Alfv\'en wave that have an average velocity with respect
to the plasma velocity. This can reduce the effective contrast in velocities experienced by the accelerating particles, resulting in steeper spectra \citep{Zirakashvili2008}. 
Although modification of the non-linear DSA model indeed predict steeper spectra, the idea that particles of different energy have experienced different shock compression ratio still stands. Thus, the expected result is still a spectrum that is steeper at low energies than at high energies, consistent with our findings.
 
Note that an altogether different explanation for spectral variation was presented by
\cite{Atoyan2000}, who suggested a two-zone model trying to explain the observed flattening in the averaged radio spectra \citep{Mezger1986}. They argued that the flattening is caused by accelerated particles escaping from bright steep-spectrum radio structures (knots, ring) into the diffuse radio `plateau'. However, this model predicts flatter spectra in the plateau and steeper at the acceleration sites. Our spatially resolved results are in contradiction with this model which shows the opposite to be true.

\subsection{Cooling break and the magnetic field}
\label{sec:cooling_break}
Both the radio-to-infrared spectral index map (Fig.~\ref{fig:radio-to-ir_sp_index}) and the index deviation map (Fig.~\ref{fig:diff_sp_index}) show regions where the flattening is minimal --- $\Delta \alpha\approx 0.05$ --- and/or the radio to infrared spectrum is still relative steep --- $\alpha_{\rm R-IR}\approx <-0.65$.

There are several possible reasons why not everywhere the spectrum flattens out to $\alpha\gtrsim -0.6$:
{\it i)} the steep spectrum and lack of flattening could be intrinsic to the particle acceleration properties that
shaped the spectrum of those regions, for example stronger magnetic-field amplification due to higher shock velocities \citep{Bell2019} as compared to more flattening regions,
or {\it ii)} it could be caused by synchrotron cooling, which has steepened the spectrum since the acceleration took place.
In the latter case the intrinsic concave curvature of the spectrum could be offset by additional steepening at higher frequencies.

Intrinsic differences in concave spectral curvature cannot be ruled out, but it is difficult to attribute that to stronger magnetic-field amplification in the steepest regions, as seen in the northern and southeastern parts of the remnant:
if anything, all indications points towards a more efficient acceleration and higher shock velocities being present in the southwest \citep[e.g.][]{Helder2008}. This is where we encounter the flattest spectra (in agreement with non-linear shock acceleration models), and where the X-ray synchrotron emission is brightest. We cannot exclude the possibility that the situation was different one or two centuries ago, when the bulk of the electrons were accelerated. 

Here we discuss the second option --- synchrotron cooling --- and the implications it may have for the acceleration history and magnetic-field strength throughout the remnant.
The synchrotron cooling effects are a function of time since the electrons have been first accelerated and the magnetic-field strength. This implies, under the assumption of steepening due to synchrotron cooling, that regions with small $\Delta \alpha$ 
have been shocked earlier, or that the magnetic-field strength is higher. 

As a reminder, relativistic electron gyrating around the magnetic field line radiates away its
energy within a cooling time scale 
\begin{equation}
    \tau _\mathrm{syn} \equiv \frac{E}{dE/dt} = \frac{9}{4} \frac{(m_{\rm e}c^2)^4}{e^4cB^2E_{\rm e}} \\
    \approx 1250 \left( \frac{E_{\rm e}}{1\mathrm{TeV}}\right)^{-1} \left( \frac{B}{100\mu \mathrm{G}}\right)^{-2} {\rm yr},
\label{eq:synchrotron_time_scale}
\end{equation}
where $m_{\rm e}$ and $e$ are the mass end electrical charge of the electron, $E_{\rm e}$ is its energy, and $B$ is the magnetic-field strength. 

Equating synchrotron loss timescale, $\tau _\mathrm{syn}$, with the age of the electron population, $t_{\mathrm{age}}$, we obtain the time-dependent energy that represents the turnover
energy in the particle spectra. Above this energy, we do not expect to detect any more particles due to synchrotron losses for a accelerated electron population with a single age.

For a given electron energy, the typical synchrotron radiation frequency is 
\citep{Ginzburg1965}
\begin{equation}
    \nu _\mathrm{syn} = 4.6 \times 10^{13} \left( \frac{B}{100\mu \mathrm{G}} \right) \left( \frac{E_e}{1\mathrm{TeV}} \right)^2 \mathrm{Hz}.
\label{eq:energy_to_frequency_age}
\end{equation} 
Combining Eq.~\ref{eq:energy_to_frequency_age} with Eq.~\ref{eq:synchrotron_time_scale} gives the typical frequency above which the synchrotron spectrum should be affected by
radiative losses (i.e. the cooling-break frequency):

\begin{equation}
    \nu _\mathrm{br} \approx 7.2 \times 10^{16} \left( \frac{B}{100\mu \mathrm{G}}\right)^{-3} \left(\frac{t_\mathrm{age}}{100~\mathrm{yr}} \right)^{-2} \mathrm{Hz}.
\label{eq:frequency_age}
\end{equation}

In a SNR, the total emission arises from a combination of synchrotron cooled
spectra with different ages for the electron populations. 
So whereas each population will have its own sharp cut-off in frequency,
the combination of different populations leads to a broken power-law spectrum,
with the break frequency corresponding to the age of the oldest electron population \cite[e.g.][]{Longair2011}, and the break frequency given by Eq.~\ref{eq:frequency_age}. 

\subsubsection{Break frequency estimation}
In order to assess whether a synchrotron cooling break
could have affected the average flattening of the spectrum we
need to make an assumption about what the asymptotic 
spectral index would be without synchrotron cooling.

\begin{figure}
	\centering
	\includegraphics[trim=0cm 0.1cm 0cm 0.5cm,clip, width=\linewidth, angle=0]{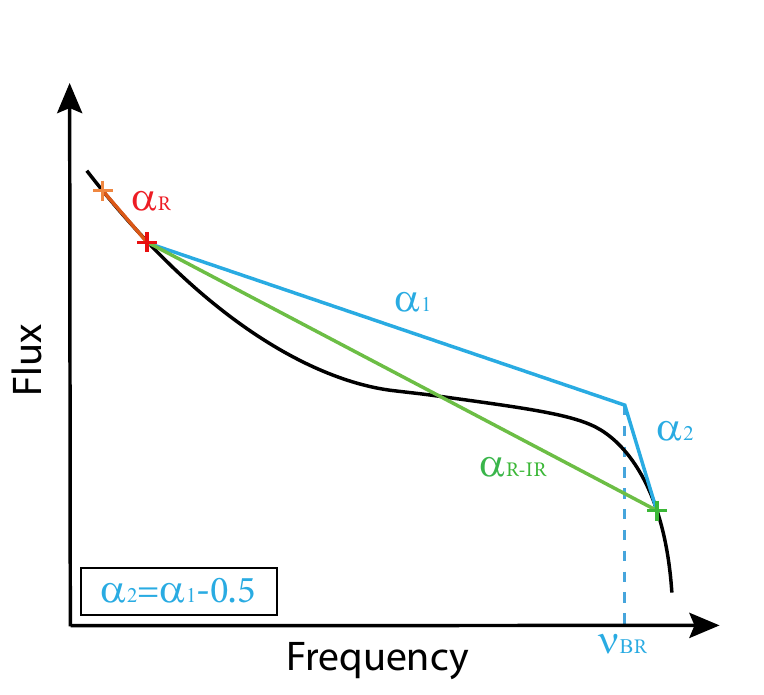}
	\caption{A sketch of what the synchrotron curve may look like, if affected by both non-linear acceleration and synchrotron cooling (solid black curve).
	The orange, red and green crosses represent the available flux data while the red and green lines are the measured radio and radio-to-infrared spectral indices.
	By making an assumption for the maximally flattening $\alpha _1$ 
	and applying a cooling break $\alpha_2 = \alpha _1 - 0.5$ (blue lines), we estimate the lower limit of frequency where the cooling break happens. 
	}
    \label{fig:fig_paper_sketch}
\end{figure}

The principle of our frequency cooling-break estimation is sketched in
Fig.~\ref{fig:fig_paper_sketch}. The three crosses represent the measured fluxes, while $\alpha _\mathrm{R}$ and $\alpha _\mathrm{R-IR}$ are the spectral indices between them. We do not know, what is the true underlying flux distribution. Nevertheless, in case of the non-linear acceleration we can generally expect the spectrum to become flatter (harder)
before ultimately being driven down by the synchrotron cooling. This is represented by the black curve in Fig.~\ref{fig:fig_paper_sketch}.

Our estimate for the frequency at which a cooling break
may be present consists of approximating the curved spectrum
by two segments with fixed power-law slopes, $\alpha_1$ --- the flattened spectrum--- and $\alpha_2$, the spectrum affected by the cooling. 
These two segments together are forced to be consistent with the average measured spectral index $\alpha_{\rm R-IR}$.
We make an assumption on how flat $\alpha _1$ can be and set it as an upper limit. We then extrapolate a line from the last observed radio point to higher frequencies. At certain frequency a cooling break occurs and changes the slope of $\alpha _1$ to the steeper $\alpha _2$. Here we make another assumption that the change between these two indices is $ \delta \alpha = \alpha _1 - \alpha _2 = 0.5 $. This assumption is motivated by calculations of continuous injection electron spectra \citep{Longair2011} 
and can be taken as a lower limit for the change in the slope. Note that larger $\delta \alpha$ leads to higher break-frequency estimations. By requiring the second slope of $\alpha _2$ to intersect with infrared flux measurement we get to the lower limit of the cooling break frequency. 

Using the available data there are two possible approaches to estimate maximum flattening of $\alpha _1$. 
The first one takes advantage of the Fig.~\ref{fig:hist_all} histogram, where we assume that the maximum flattening measured in the remnant ($\alpha _\mathrm{R-IR} \approx -0.55$) is the true flattening unaffected by synchrotron cooling everywhere. The $\alpha _\mathrm{R-IR} = -0.55$ value is close to the expected $\alpha$ from the test-particle approach ($\alpha \approx -0.50$).
Moreover, $\alpha=-0.55$ (corresponding to $p=2.1$) is consistent with, but somewhat flatter than,  the
gamma-ray spectral index of $\Gamma
\approx 2.2$ (Sect.~\ref{sec:disc_comparison_other_studie}).
We estimate the lower limit for the frequency break using 
Eq.~\ref{eq:frequency_break_delta_alpha} where $\alpha _1$ is set to be $ -0.55$ while the flux density $S$ changes depending on the position in the remnant.
The break frequency is then given by
\begin{equation}
    \nu _\mathrm{br} = \left( \frac{S_\mathrm{3.6\mu m}}{S_\mathrm{4.72GHz}} \right)^\frac{1}{\delta \alpha} \cdot \left( \frac{\nu_\mathrm{3.6\mu m}}{\nu_\mathrm{4.72GHz}} \right)^{ \frac{\alpha _1}{\delta \alpha}  } \cdot \nu_\mathrm{3.6\mu m}
    \label{eq:frequency_break_delta_alpha},
\end{equation}
with $\nu_{\rm 3.6 \mu n}=8.3\times 10^{13}$~Hz,
the central frequency of the \textit{Spitzer} image used.

The resulting map is then used as an input to Eq.~\ref{eq:frequency_age} to estimate the upper limit of the magnetic field $B$. 

An alternative assumption is to put constraints on
a possible cooling-break frequency is to assume
that the radio spectra all flatten out with a fixed
$\Delta \alpha$, rather than towards a fixed final
spectral index $\alpha_1$. Assuming that
the maximal flattening corresponds to a region in the
remnant not affected by synchrotron cooling,
we used this maximal flattening of $\Delta \alpha = 0.22$
with respect to the radio spectral index 
(see Fig.~\ref{fig:diff_sp_index}) for all regions.

This effectively relaxes the condition that every part of the remnant needs to flatten to the same $\alpha _1 = -0.55$, no matter whether it's steeper or flatter in radio, but forces the use of the lower resolution map. Using the second option $\alpha _1$ in Eq.~\ref{eq:frequency_break_delta_alpha} is calculated individually for each pixel where $\alpha _1 = \alpha _\mathrm{R} + \Delta \alpha$. 

\begin{figure*}
	\centering
	\includegraphics[width=\linewidth, angle=0]{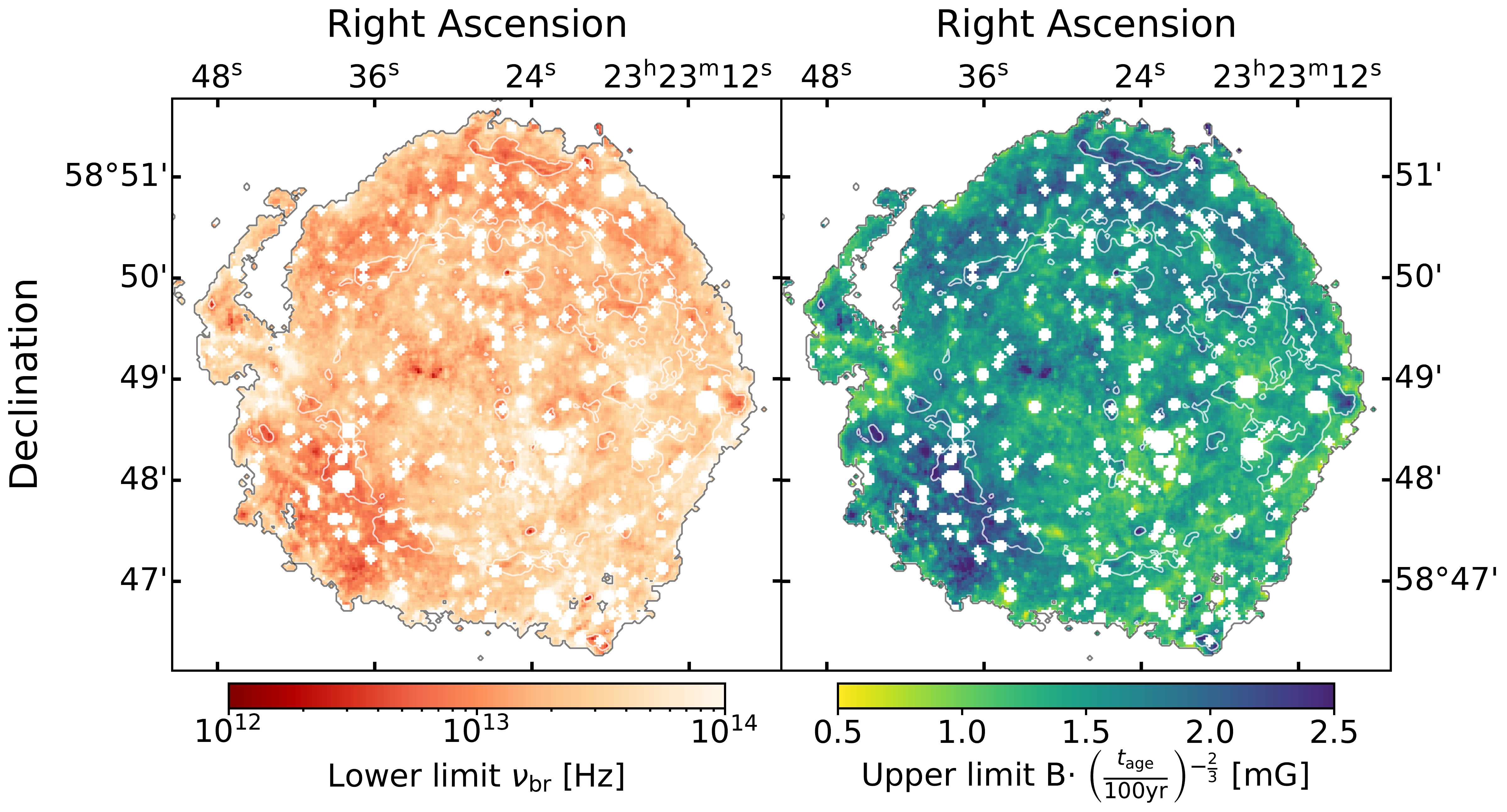}
	\caption{Estimate of the synchrotron spectrum frequency break lower limit ({left}) and magnetic field upper limit ({right}) with $\alpha _1 = -0.55$. In addition to radio contours of flux density 0.03~Jy~px$^{-1}$ we also display the outer boundary of radio-to-infrared spectral index map from Fig.~\ref{fig:radio-to-ir_sp_index}}
    \label{fig:freq_break_mag_field}
\end{figure*}

\begin{figure*}
	\centering
	\includegraphics[width=\linewidth, angle=0]{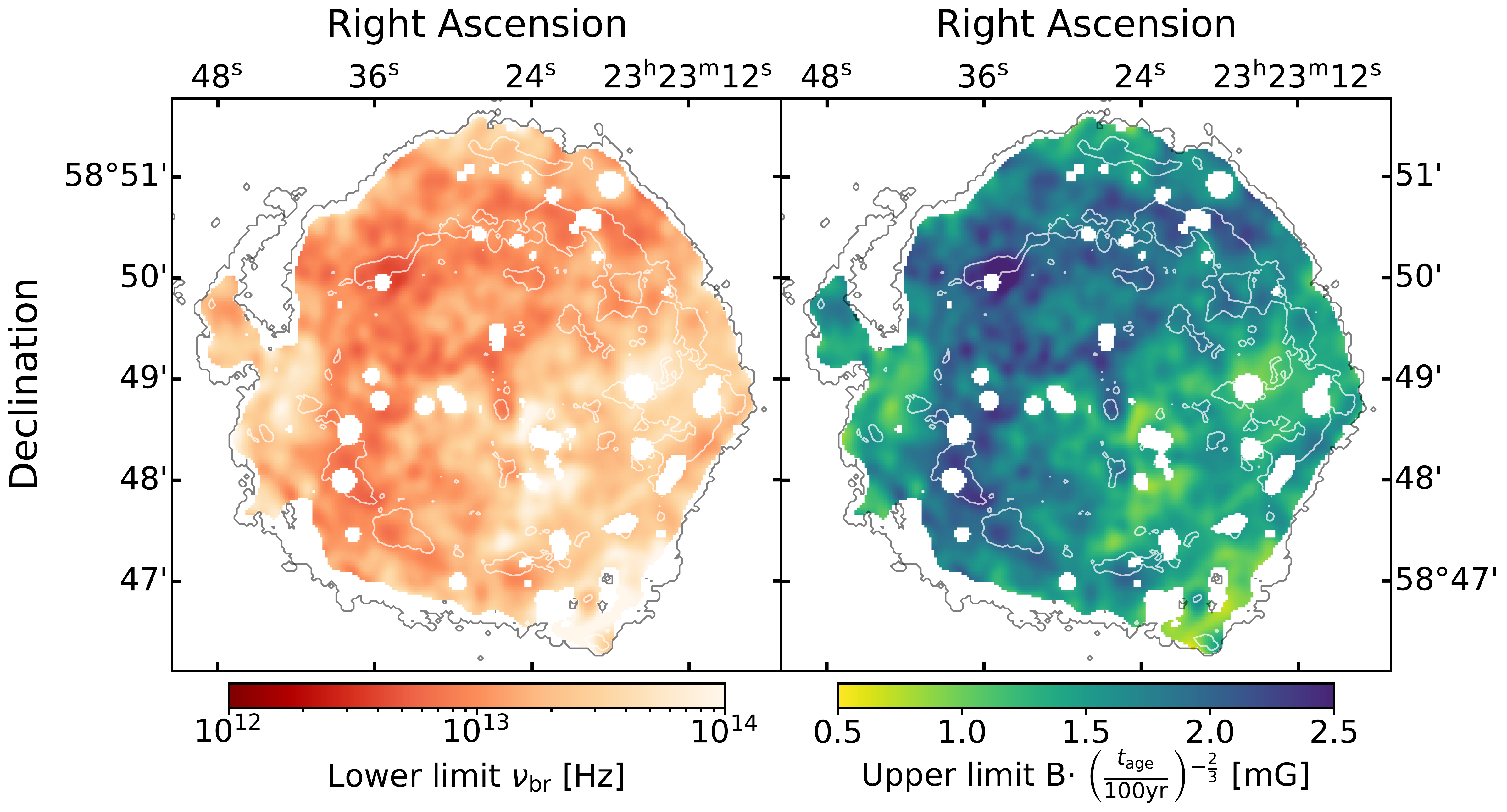}
	\caption{Estimate of the synchrotron spectrum frequency break lower limit ({left}) and magnetic field upper limit ({right}) with $\alpha _1 = \alpha _\mathrm{R} + \Delta \alpha$. In addition to radio contours of flux density 0.03~Jy~px$^{-1}$ we also display the outer boundary of radio-to-infrared spectral index map from Fig.~\ref{fig:radio-to-ir_sp_index}}
    \label{fig:freq_break_mag_field_dalpha}
\end{figure*}

Both approaches --- Fig.~\ref{fig:freq_break_mag_field} (left) and Fig.~\ref{fig:freq_break_mag_field_dalpha} (left) --- show qualitatively similar results although the detailed structures in the maps can differ.
Note that both using $\alpha_1=-0.55$ or $\Delta \alpha=0.22$ are
rather extreme values, resulting in relatively low-frequency limits for the break frequency.
In general the lower-limit estimate for the break frequency
are about an order of magnitude below the frequency of the \textit{Spitzer} observation ($\nu _\mathrm{3.6\mu m} = 8.3\times 10^{13}$~Hz). The lowest possible
cooling-break frequencies are found (not surprisingly)
in the regions of steeper spectra in the east and the north of the remnant. On the other hand in the regions where we detect strong flattening such as in the southwest, have the break frequency in the vicinity of the \textit{Spitzer} observation. In the latter case, there is effectively no need to invoke a synchrotron cooling break.

We stress here that we do not measure
a cooling-break frequency, but explore the need
for such a frequency under the assumption that
the asymptotic spectral index is $\alpha=-0.55$ throughout the remnant, or that
all spectra are concave and flatten by $\alpha=0.22$.

As such, it offers a first attempt
to map out the cooling break in the remnant.
The results suggests that, if the spectrum has cooling break
below or near the infrared map, the lowest
frequency of the cooling break are to be found in the southeast
and northeast locations.
Whether the steeper spectra there are indeed
caused by synchrotron cooling needs to be further investigated.
It is, however, difficult to directly map synchrotron radiation at lower frequencies, due the thermal emission from dust grains. So the best option would be to obtain deep near-infrared imaging. Indeed, the 2~${\rm \mu m}$ measurements by \citet{Rho2003} provides a hint that the synchrotron spectrum is generally steepening with respect to our measurements at $8.3\times 10^{13}$~Hz.

\subsubsection{Magnetic field}
Assuming that the steeper spectra as found in the southeast and
northeastern regions are caused by synchrotron cooling,
we can explore the implications for the local
magnetic-field strengths, using 
Eq.~\ref{eq:frequency_age} to transform the  limit on $\nu _\mathrm{br}$ into an upper limit of the magnetic-field strength. However, this requires the knowledge of when was the radiating plasma shocked. We used $t_\mathrm{age} = 100$~yrs in the maps presented in Fig.~\ref{fig:freq_break_mag_field} (right) and Fig.~\ref{fig:freq_break_mag_field_dalpha} (right) mostly for convenience. This places magnetic field upper limit to as high as $2-2.5$~mG in the most affected regions, while most flattening parts of the remnant in the southeast do not allow magnetic field values to rise beyond $1$~mG.
These magnetic-field upper limits are in general higher
than magnetic-field estimates in the literature, although
\citet{Atoyan2000} did invoke local magnetic-field strengths as high as 2~mG.
By using a plasma age of $100$~yr, we strike a compromise
between the maximum age, corresponding to the age of Cas A
($\sim 340$~yr) and the other extreme, namely having plasma
with electrons accelerated during the era of radio observations.

The latter option --- $\tau_{\rm age}<50$~yr ---
seems too extreme, given 
that the earliest radio-synthesis maps of Cas A ---
early 1960s \citep{Ryle1965}---
show that the overall morphology of Cas A has not
changed dramatically over the last 50~yr.
So at worst the upper limits would be 60\% higher,
if one assume the unlikely case of  $\tau_{\rm age}=50$~yr.

On the other had, it is entirely possible that a dominant fraction of the remnant's plasma was shocked in the first century after the explosion. In that case we could use the age of Cas A, which is about $\sim 340$~years. This scales down the magnetic field upper limit in Fig.~\ref{fig:freq_break_mag_field} (right) and Fig.~\ref{fig:freq_break_mag_field_dalpha} (right) by a factor 2.2, bringing the upper limits down to
0.45~mG to 1~mG.

These magnetic-field upper limits 
are consistent with other estimates of the magnetic-field strength for Cas A. 
Near the shock fronts, X-ray synchrotron observations provide estimates 
of the magnetic-field strengths of 100--500~$\mu$G \citep{Vink2003,Berezhko2004,Bamba2005,Ballet2006,Helder2012}. 
Joint modelling of the broad band spectral energy distribution of the non-thermal
radiation (radio, X-rays and gamma-rays) also suggests magnetic fields of $\gtrsim 150$~$\mu$G
\citep{Atoyan2000,Ahnen2017,Abeysekara2020}. Finally, early estimates of the overall
magnetic-field strength based on the minimum energy argument were $\sim 500$~$\mu$G \citep{Rosenberg1970}.

The fact that our location dependent magnetic-field upper-limits
are consistent with overall estimates of the magnetic fields
in Cas A adds credibility to the idea that the steeper spectra are
indeed affected by synchrotron cooling. But clearly, follow-up
studies are needed to confirm this. This is important to do,
because this is one of the few observational methods that can
inform us about the acceleration history of the relativistic
electron populations in Cas A.

\section{Conclusions}

We investigated the spatially-dependent spectral changes in synchrotron emission by comparing a high spatial resolution radio (4.72~GHz) and infrared (3.6~$\mu$m) maps of Cas A, and for the first time obtain spectral index map between these two bands. The aim was to verify earlier reports that the
the non-thermal spectrum of Cas A is flattening, consistent with predictions from non-linear DSA theory. Unlike previous studies, we obtain a spatial map of spectral indices, after correcting for interstellar extinction, and testing for potential systematic errors and also compare to radio (1.285 -- 4.64~GHz) spectral index map.

Our main findings are:

\begin{itemize}
    \item We observe flattening of the non-thermal spectrum all across the remnant with average radio to mid-infrared spectral index $\alpha _\mathrm{R-IR} = -0.61 $. This is approximately $\Delta \alpha \simeq 0.15$ flatter than the measured average spectral index in radio $\alpha _\mathrm{R} \sim -0.765$.
    
    \item The flattest indices are obtained at the locations of assumed active particle acceleration with high shock velocities, such as near the forward shock and the southwest quadrant of the remnant. These locations also coincide with the regions that are brightest in the non-thermal X-ray emission. 
    
    \item The flattening of the spectrum at high frequencies is in qualitative agreement with non-linear DSA theory. The spectral index is however still steeper than the predictions of the non-linear models without magnetic-field amplification \citep[][and references therein]{Malkov2001}.
    
    \item We asses a possibility of a cooling break being present in the remnant. We place the break frequency lower limit to $\nu _{\rm br} \sim 5\times 10^{12}$~Hz in the likely most affected regions. Considering 100~yrs as the age of the shocked plasma, this indicates allowed magnetic-field strength up to $B\sim 2-2.5$~mG in the southeast/northeast and $B\sim 1$~mG in the southwest of the remnant. This is higher than the average magnetic-field limits inferred for Cas A, but probes the magnetic-field values suggested locally by
    \cite{Atoyan2000}. Using the age of Cas A ($
    \sim 340$~yr) as the age of the plasma, puts the upper limits close
    to the magnetic-field estimates obtained with other methods.
    
    Further investigations are however still necessary and southeast/northeast regions are the best candidate locations to further investigation of the cooling break. Doing this may put
    constraints on the history of electron acceleration in Cas~A.
    
\end{itemize}

\section*{Acknowledgements}

We would like to thank Una Hwang for providing the $N_{\rm H}$ maps and Matthew Baring and Dan Milisavljevic for their insights. We are also grateful to the anonymous referee for the constructive and valuable input.
This work is based in part on observations made with the \textit{Spitzer Space Telescope}, which was operated by the Jet Propulsion Laboratory, California Institute of Technology under a contract with NASA.
The work of VD is supported by a grant from NWO graduate program/GRAPPA-PhD program. 
JVHS acknowledges support from STFC grant ST/R000824/1.

This research made use of {\sc astropy}, a community-developed core {\sc python} package for Astronomy \citep{Astropy-Collaboration:2013aa} and {\sc matplotlib} \citep{Hunter2007aa}. We further made use of {\sc SAOImage DS9} \citep{ds9_2003}, {\sc PyRAF} \citep{PyRAF2012} and the SAO/NASA Astrophysics Data System. 

\section*{Data availability}
The data underlying this article are available in the Zenodo repository, at \url{https://dx.doi.org/10.5281/zenodo.4288361}.




\bibliographystyle{mnras}
\bibliography{mnras_casa.bib} 


\newpage
\appendix

\section{Example of obtaining spectral indices per pixel using Monte Carlo simulations}\label{sec:appendix}

In order to illustrate our procedure for measuring a spectral index for a single pixel and to prove the overall robustness of our results given the
statistical and systematic errors,
we show here our procedure for a given pixel, and how it leads to a likelihood range in spectral index.

We chose a pixel in the southeast location at coordinates (Ra 23h 23m 39.1s, Dec 58 47$^\prime$ 23.8$^{\prime\prime}$) as our example pixel to provide a more complete picture of our data handling process. The location is ideal to showcase the effect of the strongest effect of background subtraction as this region provides the lowest mid-infrared flux.

{We use Monte Carlo approach in order propagate the uncertainties and obtain the result. The basis of this method is taking a value and its uncertainty of a given parameter and simulate a normal distribution around it. In our case, we produce 1D arrays of 10 000 elements. These arrays are then used as input into equations as if it would be a single value. However, as we do it on 10 000 elements we are saving the information about parameter's possible values and effectively propagate also its uncertainty.}

\subsubsection{Background subtraction}
\begin{figure*}
    \centering
    \includegraphics[width=\linewidth]{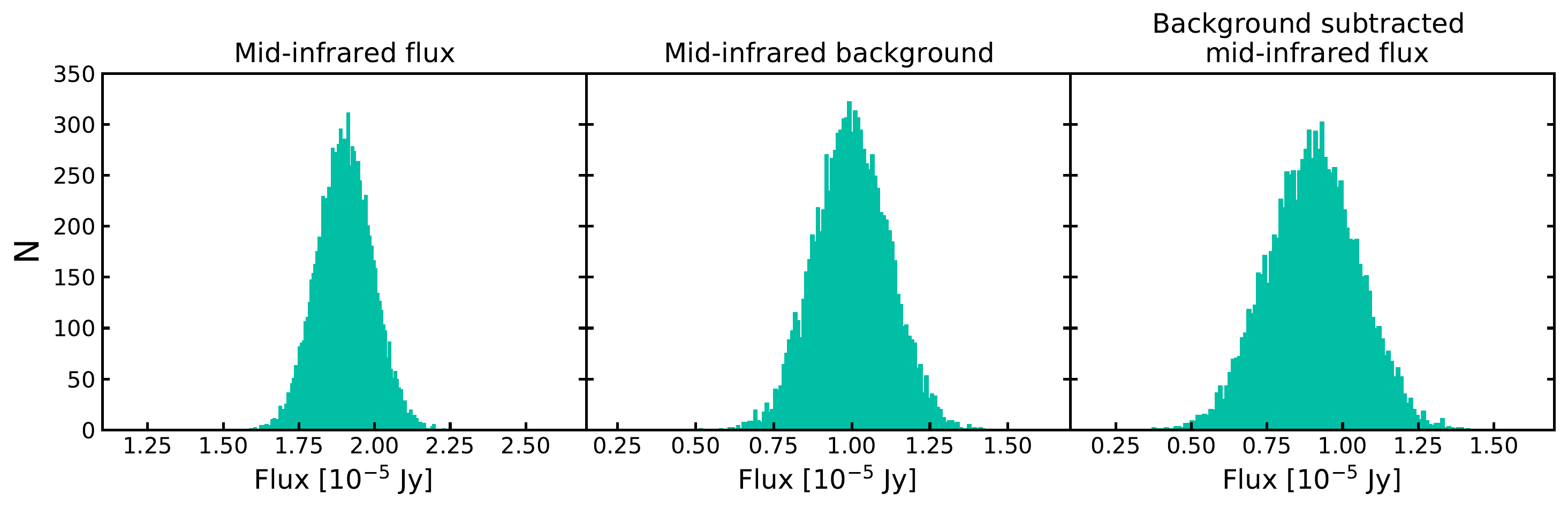}
    \caption{\textit{Spitzer} background correction process. {Left} and {middle} images show simulated normal distributions for \textit{Spitzer} data for given pixel and estimated global mid-infrared background respectively. {Right:} final background-subtracted distribution. {Colour template refers to Fig.~\ref{fig:error_diagram}.} }
    \label{figA:observed_data}
\end{figure*}

The flux in the chosen pixel is 190.1$\times 10^{-7}$~Jy~px$^{-1}$ with an uncertainty 9.1$\times 10^{-7}$~Jy~px$^{-1}$. {The mid-infrared background value was estimated to be 1.0 $\pm$ 0.15 $\times 10^{-5}$ Jy~px$^{-1}$.} Using the Monte Carlo technique we produce a normal distribution { array with 10 000 values} of both mid-infrared flux (Fig.~\ref{figA:observed_data} {left}) and estimated background (Fig.~\ref{figA:observed_data} {centre}).  

The background-subtracted distribution (Fig.~\ref{figA:observed_data} {right}) {is produced by subtraction of these two arrays and } has a Gaussian shape. This is a result that should be expected if the background correction was applied correctly. If we were over-subtracting background, the distribution would be cut from the left side and introduce inaccuracy and bias into our method. Since we chose a pixel in the low-flux southeastern region, this is additional proof that the steeper values we see there are not caused by over-subtraction.

\subsubsection{Extinction calculation}

\begin{figure*}
    \centering
    \includegraphics[width=\linewidth]{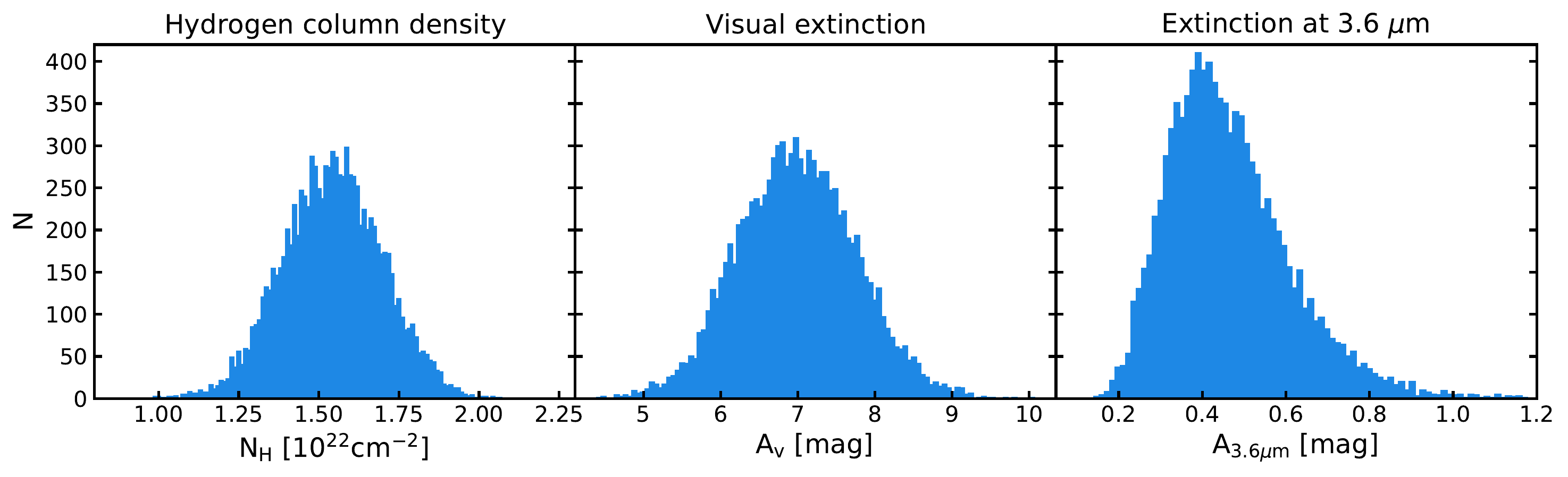}
    \caption{Extinction conversion process from hydrogen column density measured in X-rays into extinction at 3.6~$\mu$m . {Left:} distribution simulated from X-ray $N{_\mathrm{H}}$ data. {Middle:} distribution after conversion to visual magnitude $A_\mathrm{v}$ \citep[using][ eq.~1]{Guver2009}. {Right:} distribution adapted to 3.6~$\mu$m extinction \citep[using][ eq.~4]{Indebetouw2005}. {Colour template refers to Fig.~\ref{fig:error_diagram}.} }
    \label{figA:extinction_correction}
\end{figure*}

We obtain an extinction value from the X-ray measurements of hydrogen column density $N_{\mathrm{H}}$ provided by \cite{Hwang2012}. For our chosen pixel, $N_{\mathrm{H}}$ is $ 1.54 \times 10^{22}$~cm$^{-3}$. Since the uncertainty is not available we assume it to be a 10\% of the $N_{\mathrm{H}}$ value. The simulated distribution {of the 10 000 $N_{\mathrm{H}}$ values} (Fig.~\ref{figA:extinction_correction} {left}) is converted into visual extinction $A_{v}$ (Fig.~\ref{figA:extinction_correction} {centre}) {by propagating through} the $N_{\mathrm{H}} - A_{v}$ relation from eq.~1 of \cite{Guver2009} and further transformed to extinction at 3.6~$\mu$m (Fig.~\ref{figA:extinction_correction} {right}) using the extinction curve from eq.~4 of \cite{Indebetouw2005} {(See also eq.~\ref{eq_guver} and eq.~\ref{eq_indebetouw} in the subsection \ref{sec_corrections} of this paper as well for convenience)}. 

{In} this conversion we also accounted for the uncertainties {in} relations between $N_{\rm H}$, $A_{\rm V}$ and the final extinction at 3.6~$\mu$m.
The resulting extinction distribution has a longer tail in higher extinction values. This is a product of unequal uncertainties in the relation of the extinction curve and would be much harder to account for analytically. The distribution's shape is however still approximately Gaussian. {The obtained extinction distribution is then applied on mid-infrared data using the Pogson equation resulting in distribution shown in Fig.~\ref{figA:spectral_index} {left}.}

\subsubsection{Spectral index}

\begin{figure*}
    \centering
    \includegraphics[width=\linewidth]{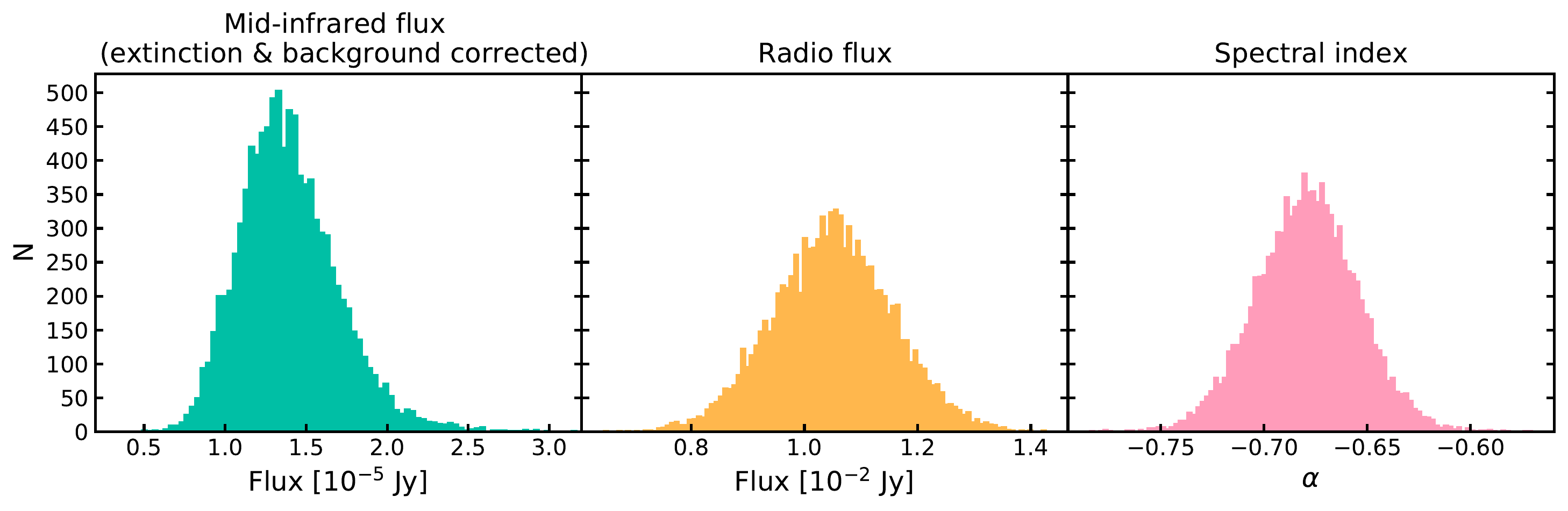}
    \caption{Obtaining the final power-law index. {Left:} background and extinction corrected distribution. {Middle:} simulated distribution based on radio data. {Right:} final spectral index distribution for the given pixel. {Colour template refers to Fig.~\ref{fig:error_diagram}.} }
    \label{figA:spectral_index}
\end{figure*}

The final step is combining the background and extinction corrected distribution of mid-infrared flux (Fig.~\ref{figA:spectral_index} {left}) with the radio data (Fig.~\ref{figA:spectral_index} {centre}) {by propagating it through the eq.~\ref{eq:spectral_index}}. The radio flux for the given pixel is $1.05\times 10^{-2}$~Jy~px$^{-1}$. As the uncertainty was not available we assumed it to be 10\% of the flux. The final result has also Gaussian shape and is shown in Fig.~\ref{eq:spectral_index} {right}.


\section{Spectral index analysis in lower resolution}
\label{sec:appendixB}
\begin{figure*}
	\includegraphics[width=\linewidth]{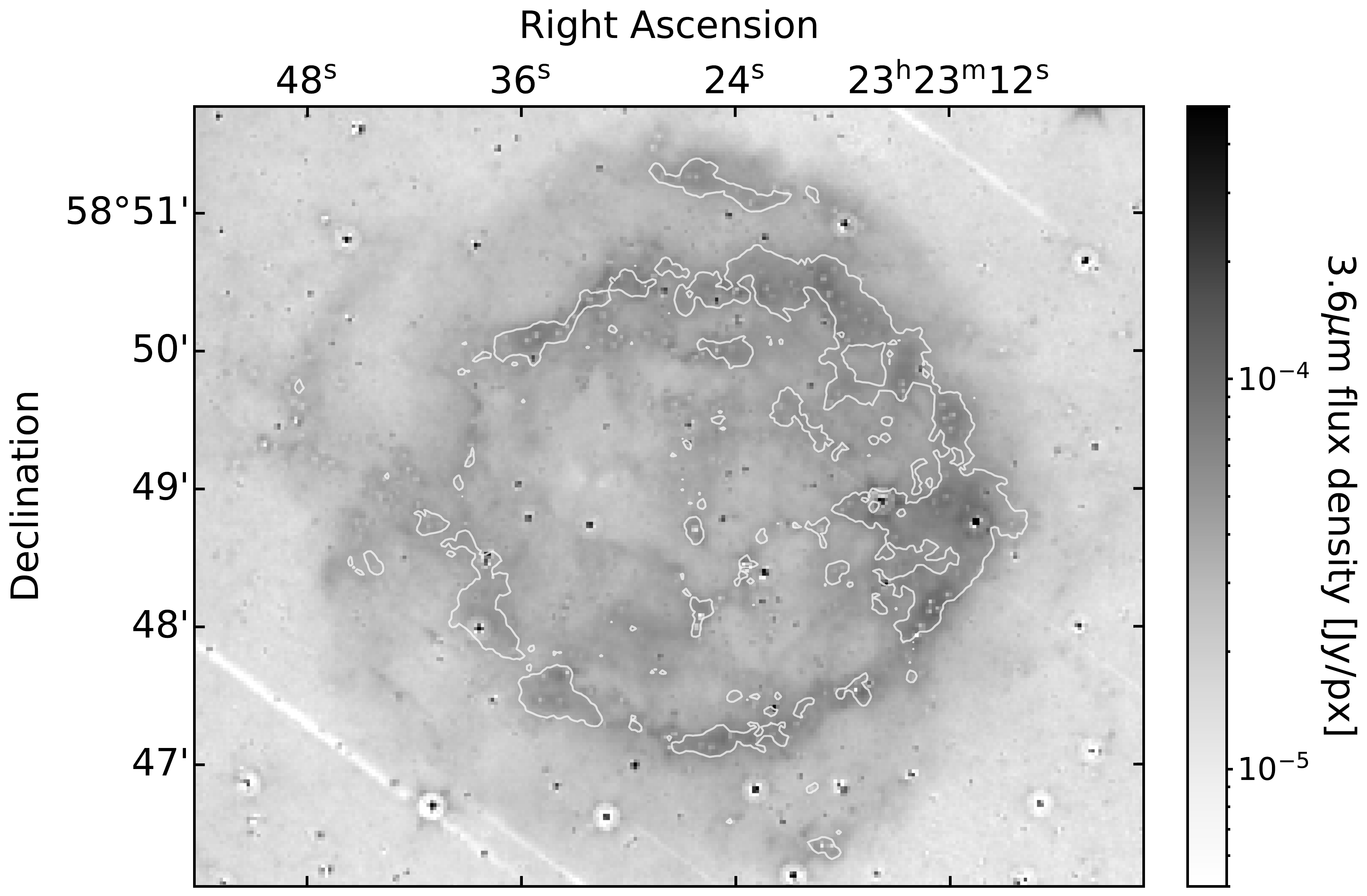}
    \caption{Point source subtracted \textit{Spitzer} 3.6 $\mu$m image from Fig.~\ref{fig:data} {Right}. White point sources show partially over-subtracted stars while black point sources were too bright or saturated to be subtracted correctly.
    Radio contours with flux density 0.03~Jy~px$^{-1}$ are displayed.}
    \label{figB:spitzer_star_sub}
\end{figure*}

\begin{figure*}
	\includegraphics[width=\linewidth]{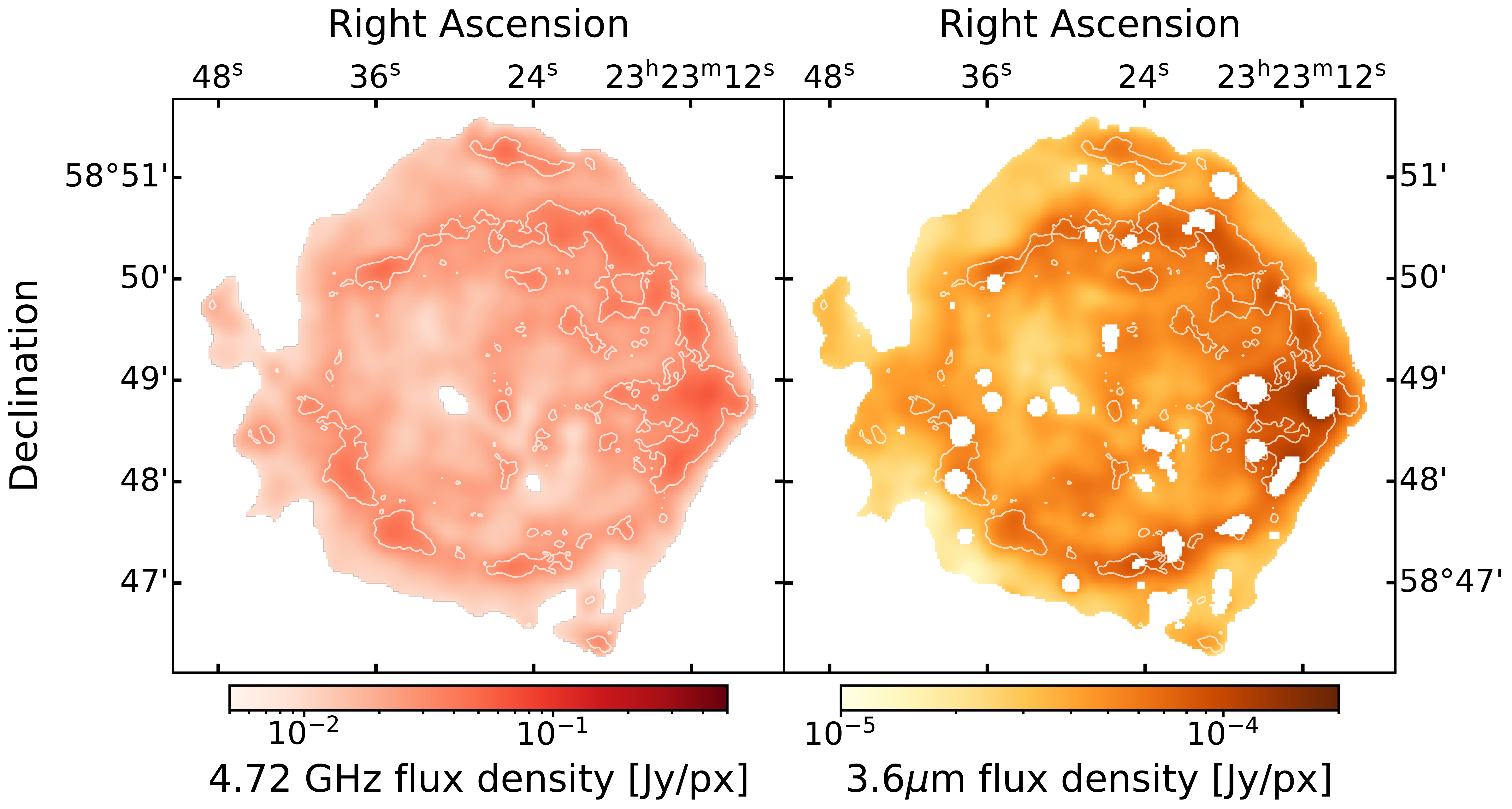}
    \caption{Flux density maps of Cas A used for construction of the $\Delta \alpha$ map in Fig.~\ref{fig:diff_sp_index}. 
    Left: Smoothed 4.72 GHz image from Fig.~\ref{fig:data} {(left)}. 
    Right: Star and background subtracted, extinction corrected and smoothed \textit{Spitzer} 3.6 $\mu$m image from Fig.~\ref{fig:data} {Right}. 
    Radio contours with flux density 0.03~Jy~px$^{-1}$ are displayed.}
    \label{figB:data_smoothed}
\end{figure*}

\begin{figure*}
	\includegraphics[width=\linewidth]{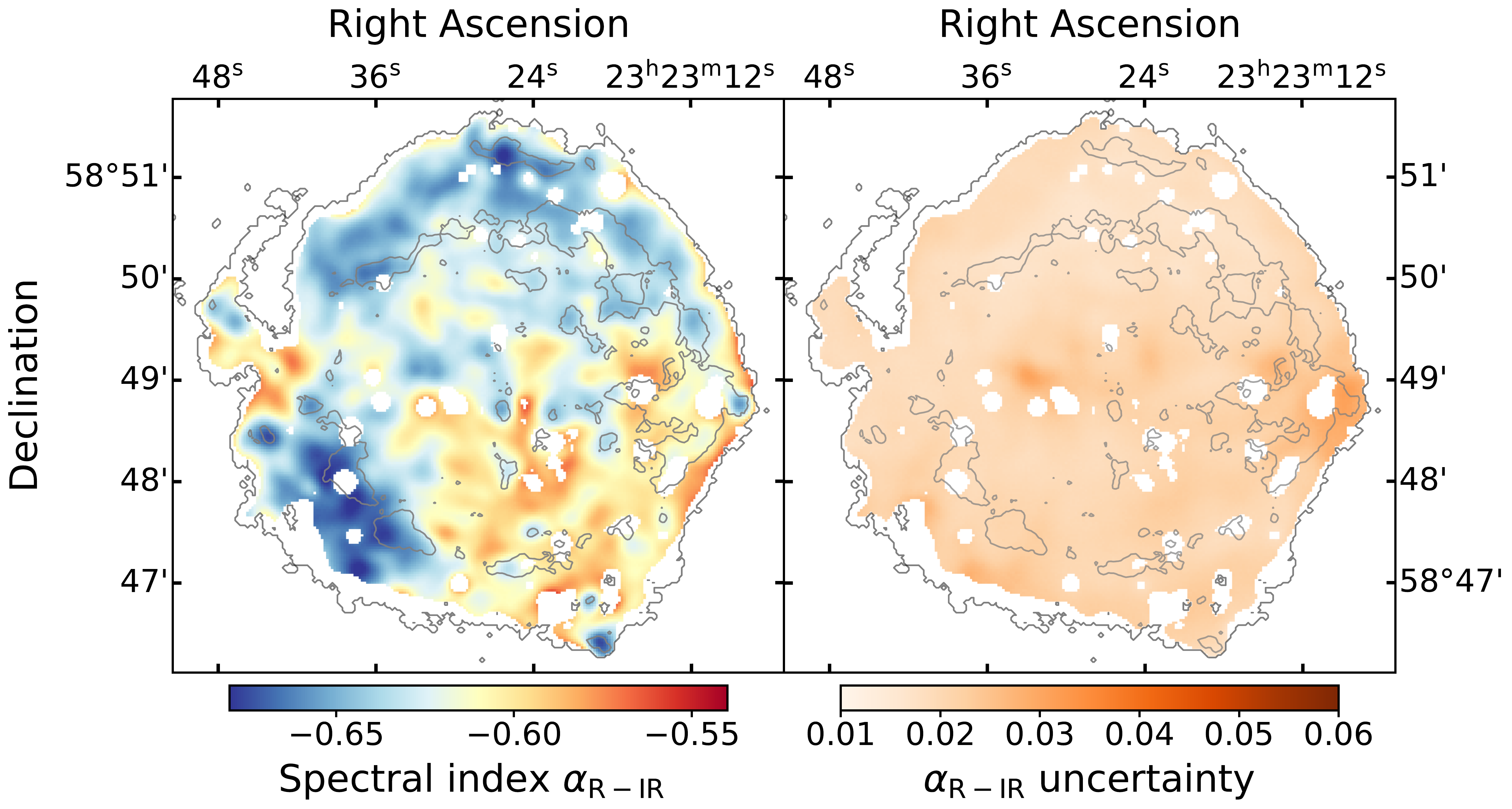}
    \caption{Radio-to-infrared spectral index map made in the 10 arcsecond spatial resolution. This map is used for construction of the spectral index difference map in Fig.~\ref{fig:diff_sp_index}. In addition to radio contours of flux density 0.03~Jy~px$^{-1}$ we also display the outer boundary of radio-to-infrared spectral index map from Fig.~\ref{fig:radio-to-ir_sp_index}}
    \label{figB:radio-to-ir_sp_index_smoothed}
\end{figure*}

In order to construct the spectral index deviation map, $\Delta \alpha$, in Fig.~\ref{fig:diff_sp_index} it was necessary to create a spectral index map at a lower resolution. We used {\sc daophot} to subtract the stars in the 3.6~$\mu$m \textit{Spitzer} data (Fig.~\ref{figB:spitzer_star_sub}) which was combined with the 4.72~GHz map smoothed to 10$^{\prime\prime}$ resolution (Fig.~\ref{figB:data_smoothed}). The final low-resolution spectral index map $\alpha _\mathrm{R-IR}$ is shown in Fig.~\ref{figB:radio-to-ir_sp_index_smoothed}. A detailed explanation of the procedure is presented in Sections \ref{sec_corrections} and \ref{sec_index_comparison}.

\bsp	
\label{lastpage}
\end{document}